# Wind Resource Assessment in Europe Using Emergy

Subodh Paudel[1,2], Massimo Santarelli[3], Viktoria Martin[4], Bruno Lacarriere[1], Olivier Le Corre[1*]

[1]Department of Energy Systems and Environment, Ecole des Mines de Nantes, GEPEA – CNRS 6144, France
[2]Department of Electrical Engineering, Eindhoven University of Technology, The Netherlands
[3]Department of Energy, Politecnico di Torino, Italy
[4]Department of Energy Technology, KTH Royal Institute of Technology, Sweden
*Corresponding Author: lecorre@mines-nantes.fr

**Abstract**

In context of increasing use of renewable sources, it is of importance to correctly evaluate the actual sustainability of their implementation. Emergy analysis is one of the possible methods useful for such an assessment. This work aims to demonstrate how the emergy approach can be used to assess the sustainability of wind energy resource in Europe. The *Emergy Index of Sustainability (EIS)* and *the Emergy Yield Ratio (EYR)* are used to analyze 90 stations of European regions for three types of wind turbines. To do so, the simplified Chou wind turbine model is used for different set of parameters as: nominal power and size of the wind turbines, and cut-in and cut-out wind speeds. Based on the calculation of the emergy indices, a mapping is proposed to identify the most appropriate locations for an implementation of wind turbines in European regions. The influence of the wind turbine type on the sustainability is also analyzed, in link with the local wind resource. Thus, it is concluded that the emergy sustainability indices are sensitive to the wind turbine design parameters (size, cut-in and cut-out wind speed).

## 1 Introduction

With the depletion of fossil fuel reserves (and its consequences on the energy prices) and a constant increasing population, new energy sources are necessary to satisfy the world energy demand. That is why renewable energy sources are more and more integrated to the countries energy mix. According to the International Energy Agency (IEA, 2012), electricity demand will grow over by 70% to almost 32000 TWh by 2035 and global renewable energy market share of electricity generation would account from 20% in 2010 to 31% in 2035. Among other renewable resources, wind power contributes to a quite significant sustainable energy market share on the global level: All wind turbines installed around the globe can produce 580 TWh/year of electricity (IEA, 2012).

In comparison to the United States and Asia, Europe has a stronger market share in the wind power potential. In 2012, Europe's wind power installed capacity was around 12.7 TW with Germany having the high-est total wind power installed capacity (30%), followed by Spain (22%), UK (8%), Italy (8%) and France (8%) (GWEC, 2012). The total electricity from wind power in Europe was 7% in 2012 (GWEC, 2012). In order to propose a sustainable electrical energy supply in the future, it is essential to conduct the wind resource assessment.

Sustainability evaluation of wind resources can be performed using different (complementary) approaches: thermo-economic analysis (energy and/or exergy calculations), life cycle assessment (which is a multi-criteria analysis product oriented,), emergy approach (a holistic approach donor side oriented)… These different assessment approaches were compared one by one (see Kharrazi et al. (2014) and/or combined (Duan et al. (2011)).Yang et al. (2012) evaluated the sustainability of a wind farm in Horqin Right Front Banner in China. They defined the so-called renewability (ratio of exergy inputs to cumulative non-renewable exergy inputs) based on the exergy accounting. The results showed that the sustainability increased with renewability. They also concluded that the renewability of wind power generation systems is higher than a hydrogen-fed steam power plant solution. For the authors, non-renewable energy inputs and environmental remediation cost constitute only a small fraction of exergy. Nguyen (2007) performed wind resources assessment based on cost-energy analysis of Vietnam using Geographic Information System (GIS). In this work, sustainability of wind resources was evaluated based on cost of electricity generation. The results showed that in 865 km2 land area, the electricity production was 3570 MW and the electricity generation cost was less than 6 US cents/kWh. Cockerill et al. (2013) performed assessment of wind resources of Northern Europe (UK, Western Ireland, Belgium, Netherlands, Germany and Denmark) on medium (150 MW) and large (400 MW) wind turbines based on energy analysis. Energy analysis was performed using GIS. The maps obtained showed the interesting areas for offshore wind farm construction and the production cost of the energy associated. Le Corre et al. (2013) performed wind resource assessment of European countries based on energy and exergy analysis. The results showed that North Sea coasts, Baltic Sea coasts, a specific coast of Mediterranean Sea and UK have higher potential areas of wind resources for electrical energy production. Wind resources assessment of Brazil was conducted by Lima et al. (2010) using Wind Atlas Analysis and Application Program (WAsP). In their work, sustainability of wind resources was evaluated in terms of payback period. Results identified the Triunfo region of Brazil as having the higher wind resources with a mean wind speed of 11.27 m/s and a payback period within 3 years.

A life cycle assessment of wind resources was performed by Demir et al. (2013) for medium (330 kW, 500 kW, 810 kW) and large (2050 kW, 3020 kW) wind turbines at Pinarbasi-Kayseri in Turkey. This assessment was done in terms of electricity production, materials used during project life cycle and energy payback period. The results delineated that environmental impacts (emissions produced during manufacturing, decommissioning and recycling) were lower for higher hub height wind turbines. Crawford (2009) performed life cycle assessments of two wind turbines (850 kW and 1650 kW) in Australia and evaluated the performance of wind resources in terms of energy yield ratio (ratio of net annual energy output for entire life span to the energy associated with manufacture, construction, installation, maintenance and replacement of wind turbines). The energy yield ratio for 850 kW and 1650 kW wind turbines was 21 and 23 respectively which signifies that both turbines were able to produce larger amounts of energy than required for their manufacture, operation and maintenance during their project span of 20 years. Also, results showed that the size of a wind turbine does not play a major role in optimizing their life cycle energy performances. A life cycle assessment was performed by Ardente et al. (2008) to measure the environmental performances of wind farms and compared with the other power generation systems. The results showed that environmental sustainability of wind resources was higher compared to the other renewable energy sources. Results also indicated that energy consumption and greenhouse gas emissions were balanced after 1-year payback period.

Emergy based sustainable indices were used by Yang et al. (2013) to compared a wind farm in China to other renewable resources. The results showed that wind power had better sustainable performance with lower environmental impact in contrast to solar thermal and photovoltaic plant systems. Dolan (2007) performed an emergy assessment of theoretical offshore wind farm on Jacksonville, Florida and compared these results to a natural gas combined-cycle unit and a coal-fired steam turbine power unit. Sustainability was compared using the classical emergy indices: *Emergy Yield Ratio* (*EYR*), *Environmental Loading Ratio* (*ELR*) and *Emergy Index of Sustainability* (*EIS*) (Ulgiati et al., 2004; Ulgiati et al., 1995; Brown and Ulgiati, 1997). The results showed that wind farm had a higher sustainability value (*EIS*) of 122 compare to 1.2 – 1.4 for the coal system and 0.6 for the natural gas system. Brown et Ulgiati (2002), assessed the environmental loading of electricity production systems using similar emergy indices (but with lower value for renewable input as *R*=7.28E17 seJ). Their results allowed a comparison among the different solutions to be done. In this ranking wind energy has the lowest Environmental Loading Ratio, followed by Geothermal, Hydro and fossil fuels. Riposo (2008) performed an emergy assessment of maple ridge wind energy facility of 195 wind turbines at New York, USA and evaluated sustainability indices in terms of electricity generation and environmental wildlife impacts. The results demonstrated the sustainability of wind energy compare to the other conventional energy.

Fig. 1 presents the complementarities between three methods of assessment (energy and exergy, LCA and emergy) in the case of sustainability assessment of wind turbines. Added to the accounting of wind turbine components contribution in the different environmental methods (dotted line), life cycle assessment includes: the extraction of raw materials (e.g. epoxy, glass fiber, cast iron, steel, copper and aluminum), materials manufacturing and land use on site (basement, tower, nacelle and rotor; concrete foundations and assembly of wind turbine), operation and maintenance to the end life span of wind turbines and wind farm disposal. As illustrated in Fig. 1, energy and exergy assessments do not consider life cycle components boundary as LCA does. In emergy analysis, direct and indirect inputs of wind turbines (including free environmental inputs) at larger time resolution are considered, whereas they are neglected or only partly accounted for in LCA.

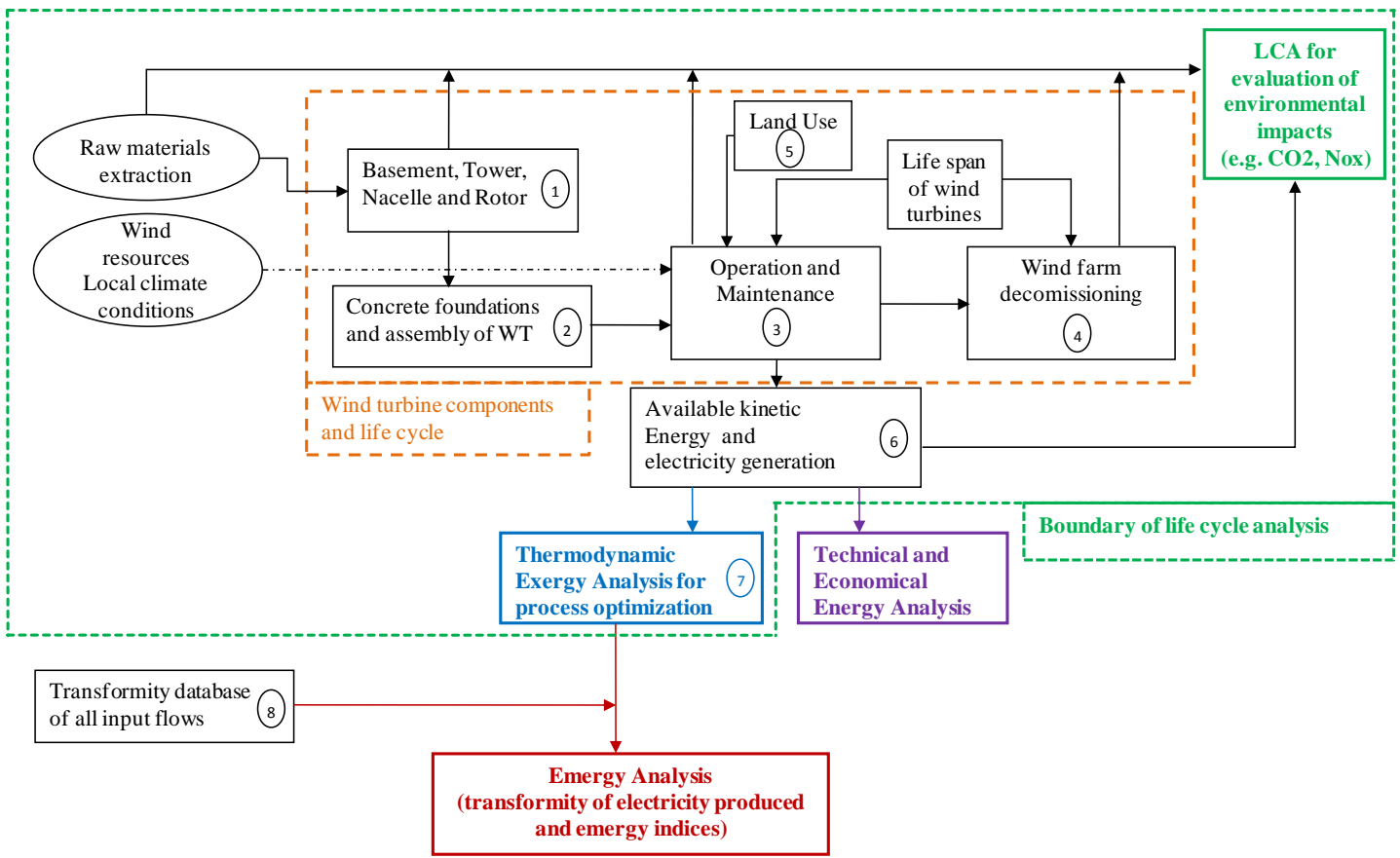

**Fig.1.** Sustainability evaluation of wind turbine assessment based on energy, exergy, LCA and emergy

In this work, emergy analysis is used to assess the sustainability of implementing wind turbine in Europe. A spatial distribution of the *transformity* of the electricity produced and *Emergy Index of Sustainability* (*EIS*) is proposed for three types of wind turbines. A sensitivity analysis of this ratio on the cut-in and cut-out wind speeds is also proposed for a selection of countries. The assessment of the sustainability of wind resources is conducted in 90 meteorological stations of European region for wind turbine sizes from 850 – 3000 kW.

## 2 Emergy Assessment

The emergy approach (originally proposed by Odum, 1996) provides useful information to assess the environmental sustainability of the exploitation of the local resources using sustainability indices. The implementation of the emergy analysis requires the inventory of all the resources necessary to produce goods, services or materials. This inventory classifies these resources into three categories which correspond to the *purchased* (*F*), the *renewable* (*R*) and the *non-renewable* (*N*) resources. In this work the *Environmental Index of Sustainability* (*EIS*) is used to assess the sustainability of windmills development in Europe. This index is defined as the ratio of the *Emergy Yield Ratio* (*EYR*) to the *Environmental Loading Ratio* (*ELR*) (Brown et Ulgiati, 2002). For these authors *EYR* measures the availability of the local resources and can be used to estimate the potential of local resources contribution on economy to society. It is defined as the ratio of the *yield emergy* (*Y=F+N+R*) to the purchased *emergy* (*F*) (*EYR=F+N+R/F*). *ELR* measures the thermodynamic distance of the system from natural ecosystem. It is the ratio of *non-renewable emergy* (*N*) and *purchased emergy* (*F*) to the *renewable emergy* (*R*) (*ELR=(F+N)/R*). Thus, *Emergy Index of Sustainability* (*EIS*) measures the sustainability of the system per unit of environmental loading.

For wind turbines, *purchased resources* (*F*) correspond to materials for basement, tower, nacelle and rotor, see block (1) of Fig. 1; concrete foundations and assembly of wind turbine plant, see block (2) of Fig. 1; services and operation and maintenance of labor, see block (3) of Fig. 1; and labor and fuel decommissioning, see block (4) of Fig. 1. The *renewable resource* (*R*) corresponds to the available kinetic energy from wind as represented in block (6) of Fig. 1 and land used by wind turbines, see block (5) of Fig. 1. It requires an estimation of the wind emergy according to the location of the wind turbine. So an assessment of wind resources is performed, using hourly wind speed data for the year 2012 (8760 data/year) from the meteorological database (DOE, 2012), taking into account the adjustment of wind speed depending on the height. Then, hourly kinetic energy available from this wind resource is calculated, taking into account the cut-in and cut-off wind speeds. The electrical energy produced is calculated using simplified Chou wind turbine model (Chou and Korotis, 1981). Then cumulated kinetic energy over the year and the corresponding electricity produced are used to draw the maps of *EIS*. All energy/exergy calculations are detailed in appendix A. Based on these local wind characteristics, total emergy is calculated (including available kinetic energy, materials, labor, fuel and land) and the transformity of the electricity produced by the wind turbines is calculated for each meteorological stations.

## 3 Case study

The case study corresponds to the 90 stations of 26 Countries of European region (Austria, Belarus, Belgium, Bulgaria, Czech, Denmark, Finland, France, Germany, Greece, Hungary, Ireland, Italy, Netherlands, Norway, Poland, Portugal, Romania, Serbia, Slovakia, Spain, Sweden, Switzerland, Turkey, Ukraine and United Kingdom) as shown in Fig. 2. Hourly wind speed values are associated to a reference height of 5 meter (see $z_0$ in Eq. 1 in appendix A). Annual average wind speed of meteorological stations is shown in Fig. 3 (in this figure the wind speeds correspond to a height of 5 meters). It is clear that a larger part of the area is dominated by average wind speed of 6-7 m/s. Only few areas have average wind speed lower than 5 m/s and greater than 9 m/s. Meteorological stations at Copenhagen, Brest, Istanbul, Hemsby, Leuchars and Amster-dam have an average annual wind speed higher than 9 m/s. Cities like Venice, Oslo, Geneva, Insbruck and Cluj have an average annual wind speed lower than 5 m/s.

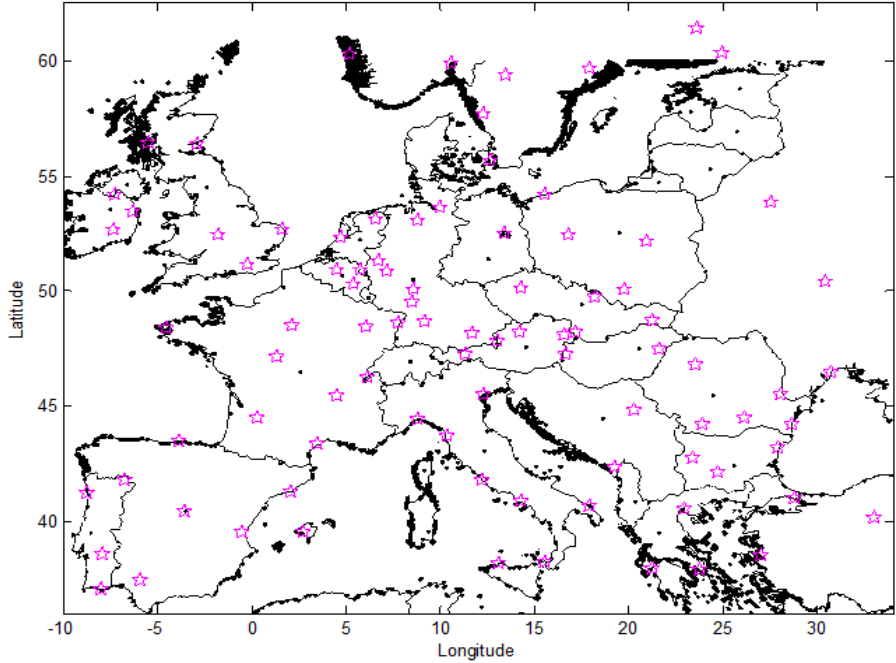

**Fig.2.** Location of 90 meteorological stations

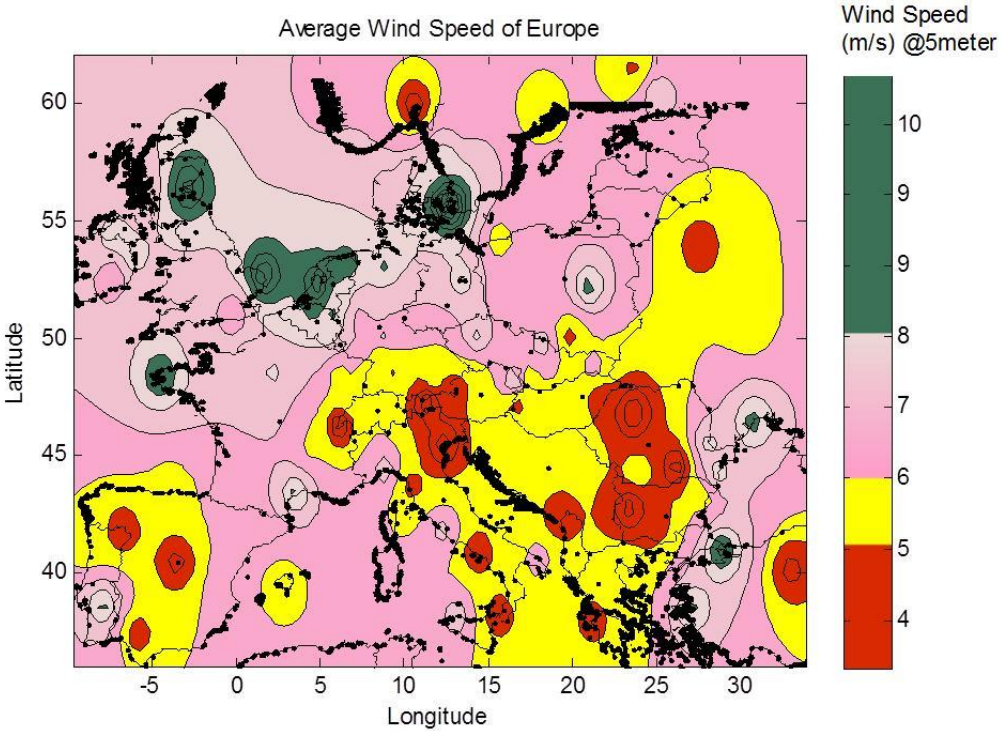

**Fig.3.** Annual average wind speed distribution for each meteorological station of European countries

Three types of wind turbines are studied in this work (850 kW, 1650 kW and 3000kW). The corresponding hub heights (*z* in Eq. 1 in appendix A) are at 68, 70 and 80 meters for the 850 kW, 1650 kW and 3000 kW wind turbines respectively. Technical specifications of wind turbines used in the in Eq. 9 (see in Appendix A) are summarized in Table 1 (Wind power, 2013).

**Table 1** Technical specifications of wind turbines (Wind power, 2013)

| Specifications | Products | | |
| --- | --- | --- | --- |
| | V52/850 | V82/1650 | V90/3000 |
| Nominal Power (kW) | 850 | 1650 | 3000 |
| Rotor Diameter (m) | 52 | 82 | 90 |
| Cut in wind speed (m/s) | 4 | 3.5 | 4 |
| Nominal wind speed (m/s) | 16 | 13 | 16 |
| Maximum wind speed (m/s) | 25 | 25 | 25 |
| Minimum hub height (m) | 65 | 59 | 65 |
| Maximum hub height (m) | 86 | 78 | 105 |
| Number of blades | 3 | 3 | 3 |

The wind turbines are constituted by rotor (hub and blades), nacelle (gearbox, generator, and brakes), tower and foundation components which are considered for the emergy assessment. Material and components of wind turbines are given in Table 2 (Crawford, 2009; Riposo, 2008). The two references used are homogeneous concerning the materials and system components. All these elements are supposed to have a same lifespan of 20 years which is also the project lifespan. In this analysis services, labor (O&M), fuel (decommissioning), labor (decommissioning) and land approximation are estimated by linear relationship from Riposo (2008). Transportation of wind turbines components to wind plant site is not considered nor the energy use to recycle materials and the energy saved through the recovery of materials at the end of turbine's life (Crawford, 2009). Cost analysis and losses associated with the grid infrastructure are not considered.

## 4 Results and discussion

Emergy values for the different wind turbines system components for the whole lifetime assessment are presented in the Table 3. It can be seen that service, labor (O&M), concrete and steel have higher emergy than other components of wind turbine system. In this table all the components are accounted for *purchased materials* (*F*). There is no local *non-renewable resources* (*N=0*).

Table 4 illustrates the emergy calculations using wind resources at Copenhagen (Denmark) as an example. Based on the local kinetic energy and the transformity of wind from Riposo (2008), the electrical output is determined. From these values the *yield emergy* (*Y*), the *renewable emergy* (*R*), the *non-renewable emergy* (*N*), the *purchased emergy* (*F*), the *Emergy Yield Ratio* (*EYR*), the *Emergy Loading Ratio* (*ELR*) and the *Emergy Index of Sustainability* (*EIS*) are calculated for the three types of wind turbines. The results given in this table for Copenhagen are consistent with those proposed by Brown et Ulgiati (2002). In their work, the authors studied different systems of electricity production among which the electricity produced from wind resources. For a 2.5MW wind turbine they found an *EYR*=7.47 (where in this study the values vary between 3.5 and 4.7), an *ELR*=0.15 (between 0.3 and 0.4 in this work) and an *EIS*=48.3 (between 8.6 and 17.8 for the present study). The differences with these references can be explained by the difference on the wind emergy as the Table 4 is only for the case of Copenhagen.

**Table 2** Materials and system components of wind turbines (Crawford, 2009; Riposo, 2008)

| Component | Item | 850 kW WT (Crawford, 2009) | | | 1650 kW WT (Riposo, 2008) | | | 3000 kW WT (Crawford, 2009) | | |
|---|---|---|---|---|---|---|---|---|---|---|
| | | Weight (g) | Materials | Others | Weight (g) | Materials | Others | Weight (g) | Materials | Others |
| Foundation | Reinforced concrete | 495 t | 480 t concrete | | 832 t | 805 t concrete | | 1176 t | 1140 t concrete | |
| | | | 15 t steel | | | 27 t seel | | | 36 t steel | |
| Tower | Painted Steel | 70 t | 69.07 t steel | | 128.2 t | 126 t steel | | 160 t | 158.76 t steel | |
| | | | 0.93 t paint | | | 2.2 t paint | | | 1.24 t paint | |
| | | | | | | 2.6 t aluminium | | | | |
| Nacelle | Bedplate/frame | 3.35 t | 3.35 t steel | | | | | 13 t | 13 t steel | |
| | Cover | 2.41 t | 2.41 t steel | | | | | 9.33 t | 9.33 t steel | |
| | Generator | 1.84 t | 1.47 t steel | | | | | 7.14 t | 5.71 t steel | |
| | | | 0.37 t copper | | | | | | 1.43 t copper | |
| | Main shaft | 4.21 t | 4.21 t steel | | | | | | | |
| | Brake system | 0.26 t | 0.26 t steel | | | | | 1.02 t | 1.02 t steel | |
| | Hydraulics | 0.26 t | 0.26 t steel | | | | | | | |
| | Gearbox | 6.2 t | 6.08 t steel | | | | | 24.06 t | 23.58 t steel | |
| | | | 0.062 t copper | | | | | | 0.241 t copper | |
| | | | 0.062 t aluminium | | | | | | 0.241 t aluminium | |
| | Cables | 0.42 t | 0.18 t aluminium | | | 9.79 t aluminium | | 1.63 t | 0.69 t aluminium | |
| | | | 0.24 t copper | | | | | | 0.94 t copper | |
| | | | | | | 9.74 t plastic | | | | |
| | Revolving system | 1 t | 1 t steel | | | | | 3.87 t | 3.87 t steel | |
| | Crane | 0.26 t | 0.26 t steel | | | | | 1.02 t | 1.02 t steel | |
| | Transformer/sensors | 1.79t | 0.894 t steel | | | 0.785 t steel | | 6.93 t | 3.47 t steel | |
| | | | 0.357 t copper | | | | | | 1.38 t aluminium | |
| | | | 0.357 t aluminium | | | | | | 1.38 t aluminium | |
| | | | 0.18 t plastic | | | 0.033 t plastic | | | 0.7 t plastic | |
| | Aggregate steel | | | | | 27.1 t steel | | | | |
| | Aggregate copper | | | | | 1.52 t copper | | | | |
| | Aggregate plastic | | | | | 1 t plastic | | | | |
| Rotor | Hub | 4.8 t | 4.8 t steel | | 4.2 t | | | 19.2 t | 19.2 t steel | |
| | Blades | 5.02 t | 3.01 t fibre glass | | | 1.8 t fibre glass | | 20.07 t | 12.04 t fibre glass | |
| | | | 2.01 epoxy | | | | | | 8.03 t epoxy | |
| | Bolts | 0.18t | 0.18 t steel | | | | | 0.73 t | 0.73 t steel | |
| Labor | Services | | | 9.25E+05 $ | | | 1.79E+06 $ | | | 3.26E+06 $ |
| & Fuel | Labor (O&M) | | | 2.11 man yrs | | | 4.10 man yrs | | | 7.46 man yrs |
| | Labor (Decommissioning) | | | 1.59 man yrs | | | 3.08 man yrs | | | 5.59 man yrs |
| | Fuel (Decommissioning) | | | 4.94E+05 g | | | 9.59E+05 g | | | 1.74E+06 g |
| Land Appropriation | | | | 2123 meter sq | | | 5281 meter sq | | | 6361 meter sq |

**Table 3**: Emergy accounting for wind turbines system components

| Components of WT | Unit | Unit Emergy Value | Ref. | 850 kW WT | | 1650 kW WT | | 3000 kW WT | |
|---|---|---|---|---|---|---|---|---|---|
| | | | | Raw Amount | Emergy (seJ) | Raw Amount | Emergy (seJ) | Raw Amount | Emergy (seJ) |
| Materials (F) | | | | | | | | | |
| Concrete | g | 2,41E+09 | [a] | 4,80E+08 | 1,16E+18 | 8,05E+08 | 1,94E+18 | 1,14E+09 | 2,75E+18 |
| Steel | g | 5,60E+09 | [a] | 1,09E+08 | 6,11E+17 | 1,85E+08 | 1,04E+18 | 2,76E+08 | 1,54E+18 |
| Fiberglass and composites | g | 1,32E+10 | [a] | 3,01E+06 | 3,97E+16 | 1,80E+06 | 2,38E+16 | 1,20E+07 | 1,59E+17 |
| alumium | g | 2,13E+10 | [a] | 5,99E+05 | 1,28E+16 | 1,29E+07 | 2,74E+17 | 2,01E+07 | 4,27E+17 |
| copper | g | 1,14E+11 | [a] | 1,03E+06 | 1,17E+17 | 1,52E+06 | 1,73E+17 | 1,67E+06 | 1,91E+17 |
| plastic | g | 6,37E+08 | [a] | 2,19E+06 | 1,40E+15 | 1,08E+07 | 6,86E+15 | 8,73E+06 | 5,56E+15 |
| paint | g | 2,52E+10 | [a] | 9,30E+05 | 2,34E+16 | 2,20E+06 | 5,53E+16 | 1,24E+06 | 3,12E+16 |
| Labor and Services (F) | | | | | | | | | |
| services | $ | 1,34E+12 | [a] | 9,25E+05 | 1,24E+18 | 1,79E+06 | 2,41E+18 | 3,26E+06 | 4,38E+18 |
| labor (O&M) | man yrs | 4,70E+17 | [a] | 2,11E+00 | 9,92E+17 | 4,10E+00 | 1,93E+18 | 7,46E+00 | 3,50E+18 |
| labor (decommissionning) | man yrs | 1,58E+17 | [a] | 1,59E+00 | 2,50E+17 | 3,08E+00 | 4,85E+17 | 5,59E+00 | 8,82E+17 |
| fuel (decommissionning) | g | 4,95E+09 | [a] | 4,94E+05 | 2,44E+15 | 9,59E+05 | 4,74E+15 | 1,74E+06 | 8,63E+15 |
| Land Appropriation (R) | meter sq | 1,34E+11 | [a] | 2124 | 2,85E+14 | 5281 | 7,08E+14 | 6362 | 8,53E+14 |

**Ref.: [a]: Riposo (2008)**

**Table 4** Emergy indices calculation (example of Copenhagen, Denmark)

| Wind Turbine kW | Unit Emergy Value (seJ/unit) | Ref. | Kinetic Energy, KE (J) | F (seJ) | R=(KE*τ) (seJ) | Y (seJ) | EYR= Y/F | ELR= (F+N)/R | EIS= EYR/ELR |
|---|---|---|---|---|---|---|---|---|---|
| 850kW | 4,19E+03 | [a] | 2,62E+15 | 4,45E+18 | 1,10E+19 | 1,54E+19 | 3,5 | 0,4 | 8,6 |
| 1650 kW | 4,19E+03 | [a] | 7,44E+15 | 8,34E+18 | 3,12E+19 | 3,96E+19 | 4,7 | 0,3 | 17,8 |
| 3000 kW | 4,19E+03 | [a] | 1,00E+16 | 1,39E+19 | 4,19E+19 | 5,58E+19 | 4,0 | 0,3 | 12,2 |

**Ref.: [a]: Riposo (2008)**

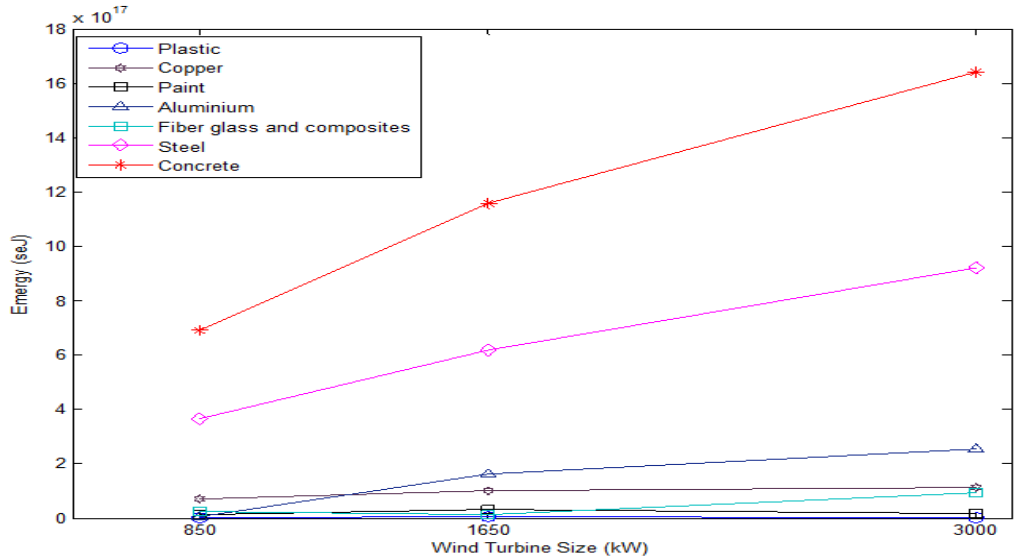

**Fig.4.** Variation of emergy of materials with wind turbines (input data from Crawford (2009) and Riposo (2008))

It can be noted that the three indicators are close to the ones proposed by Brown et Ulgiati (2002) for other renewable energy for electricity production (e.g. $EIS_{Geothermal}$=11.048 and $EIS_{Hydro}$=16.903). It can be seen in the table that *EIS* does not have a monotonic variation with the size of the wind turbine (increasing from the 850kW wind turbine to the 1650kW one and decreasing from the 1650kW wind turbine to the 3000kW one). This can be explained by a higher need of purchased inputs (more materials are required) to oversized the wind turbine compare to the renewable emergy. Indeed the limitation of the wind that can be transformed into electricity is a function of the local distribution of the wind, as it will be discussed later in this section. It is also the consequence of the second type of wind turbine characteristics (1650kW) which has lower values nominal wind speed and cut-in wind speed. The results given in Table 4 are also consistent with the definitions of the indicators as a small value for *ELR* means a low stress on the environment and the higher *EIS* is, the more sustainable the proposed solution is. Thus, the emergy indices *EYR* and *EIS* can be used to compare different locations concerning the sustainable potential they offer to use the wind resource for electricity production.

    Fig. 4 shows the emergy of materials (concrete, steel, fiber glass and composites, aluminum, copper, plastic and paint) for different wind turbines. It can be seen the almost linear relationship between the emergy and the installed power for most of the material and the low influence of fiberglass and composites, plastic and paints. It can also be shown the quasi-linearity of the electricity produced as function of the nominal power of the turbines installed (graph not presented). Based on this, the interpolations done for the electricity produced and the emergy of the wind turbines enables to plot the curve of the transformity of electricity produced as a function of the wind turbine nominal power. The result is given in Fig. 5 (through the example of Copenhagen) which shows the expected hyperbolic shape of this function. A similar calculation of the transformity of the electricity produced is done for all the 90 meteorological stations, for the three types of wind turbines. These values are given in the maps of Fig. 6 to Fig. 8.

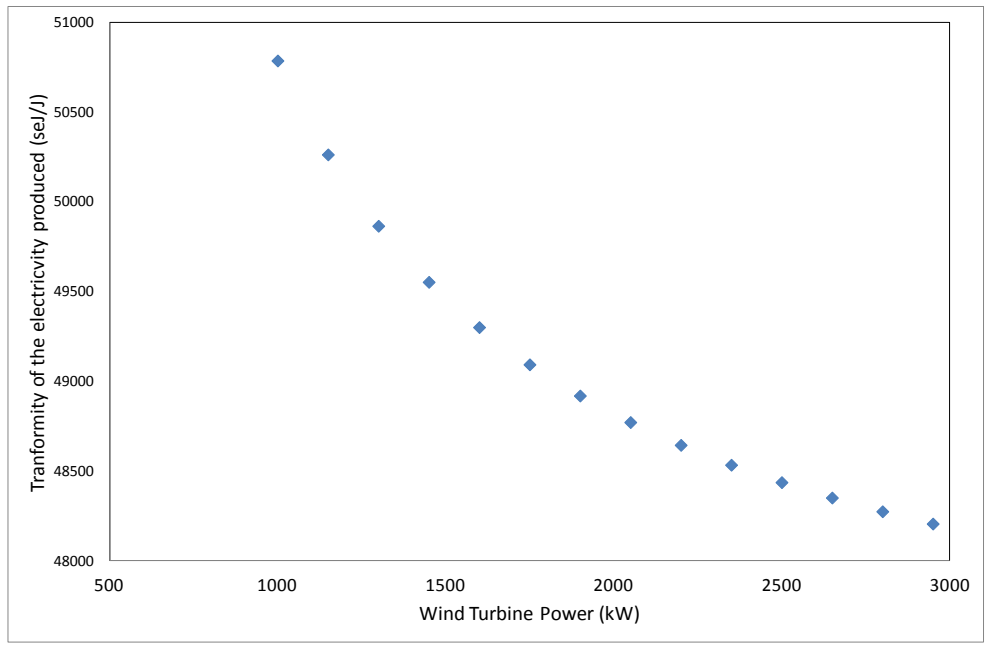

**Fig. 5.** Transformity of the electricity produced Vs installed power of wind turbines (example of Copenhagen)

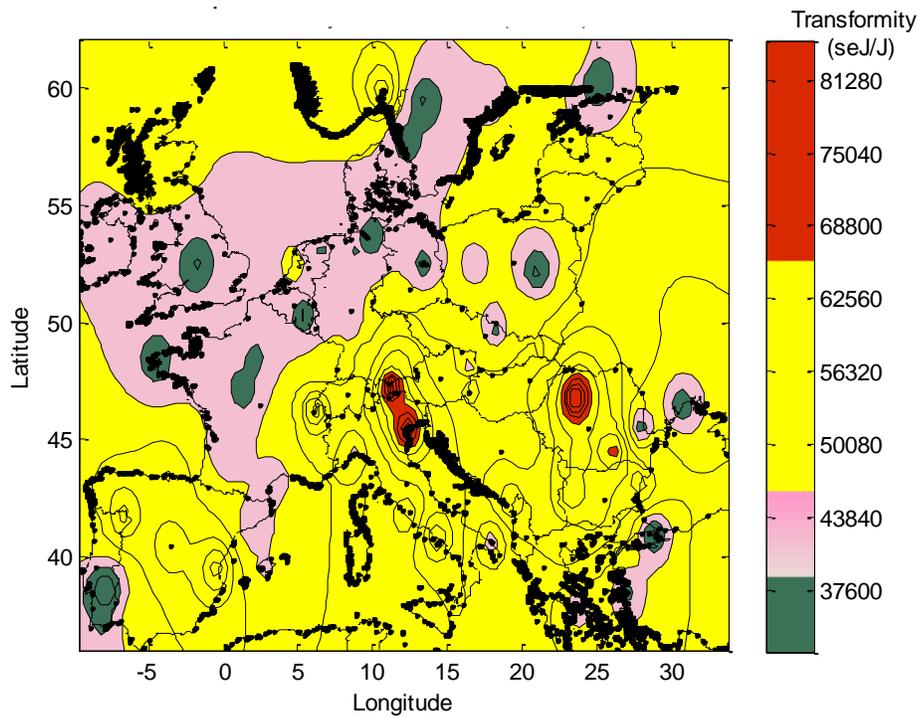

**Fig. 6.** Transformity of the electricity produced (installed power: 850 kW)

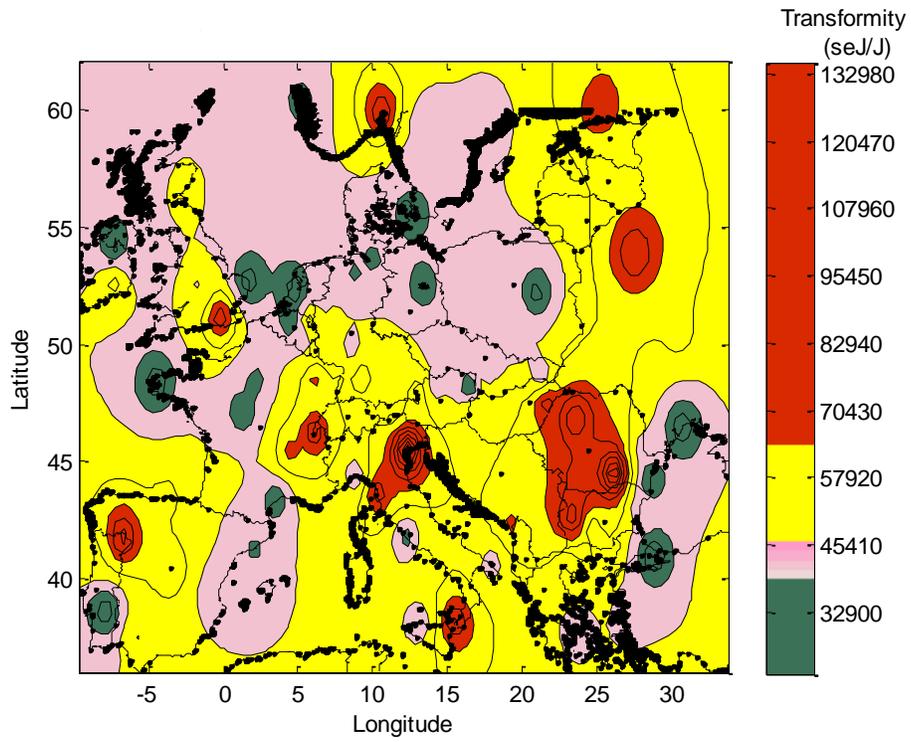

**Fig. 7.** Transformity of the electricity produced (installed power: 1650 kW)

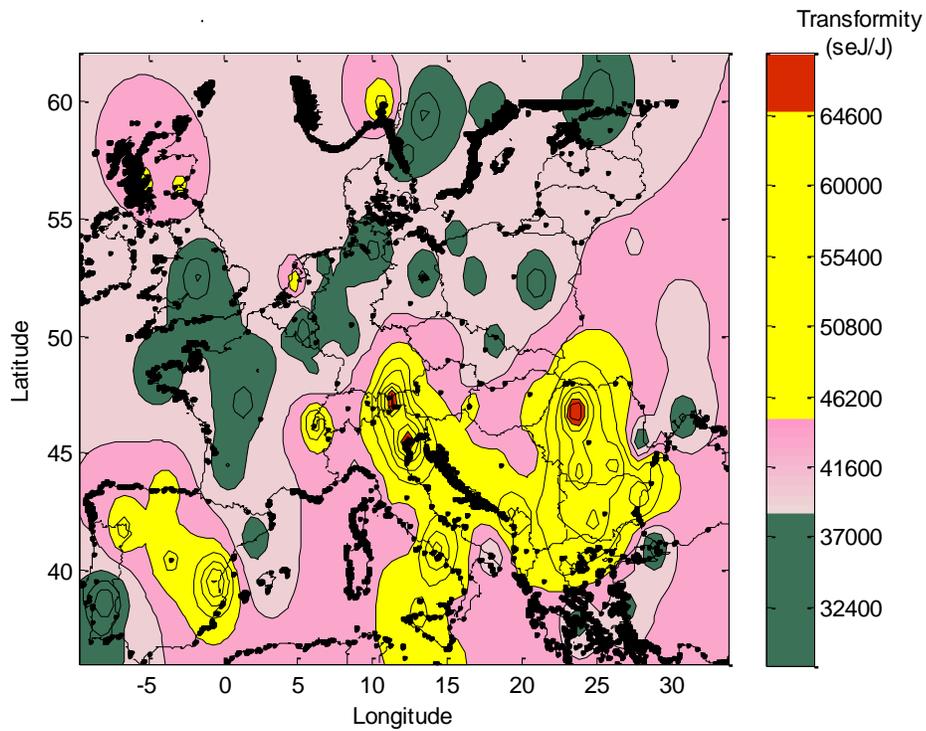

**Fig. 8.** Transformity of the electricity produced (installed power: 3000 kW)

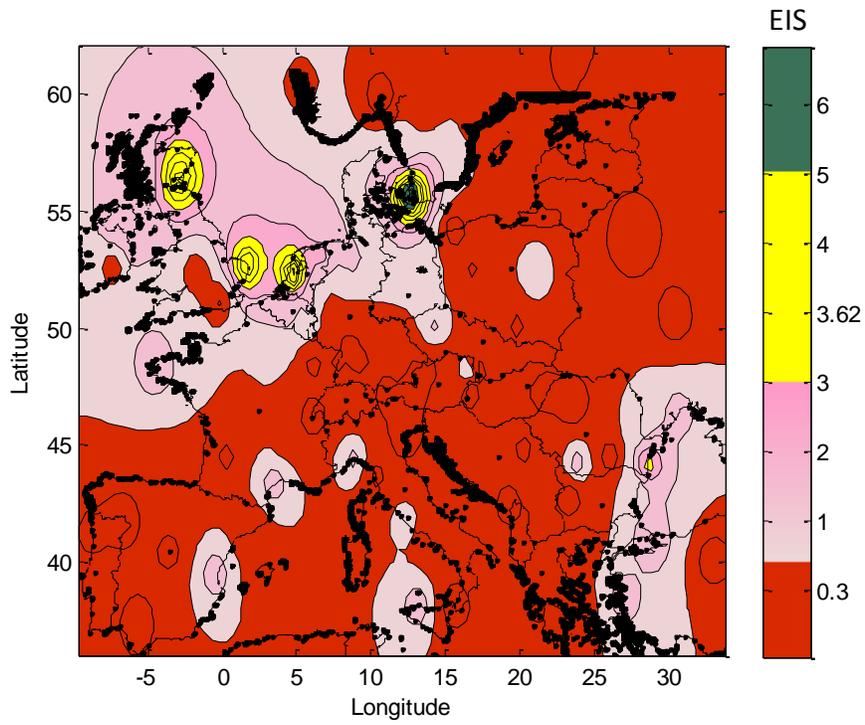

**Fig. 9.** Emergy Index of Sustainability, *EIS* (Installed power: 850kW)

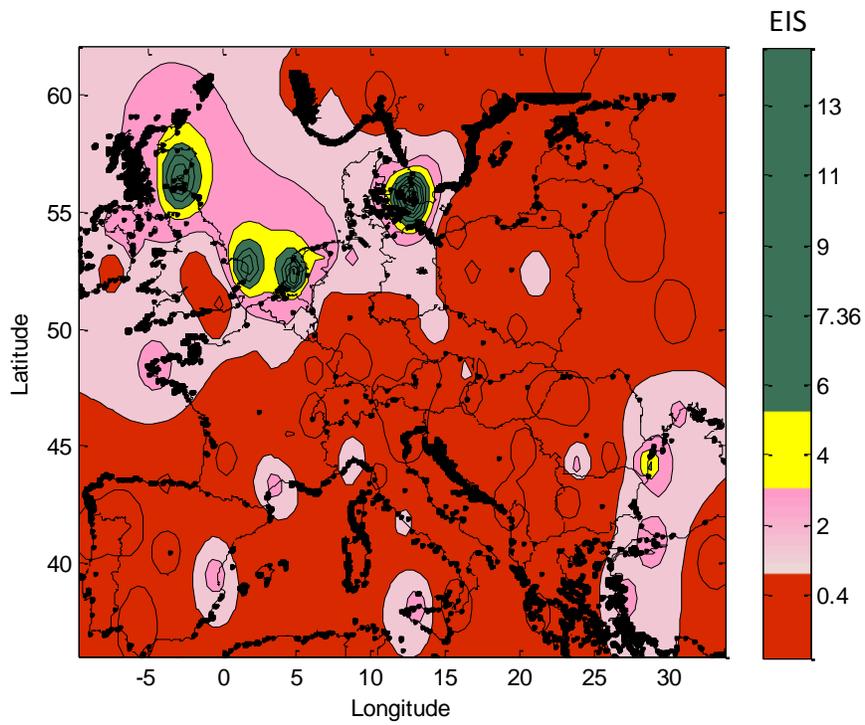

**Fig. 10.** Emergy Index of Sustainability, *EIS* (Installed power: 1650kW)

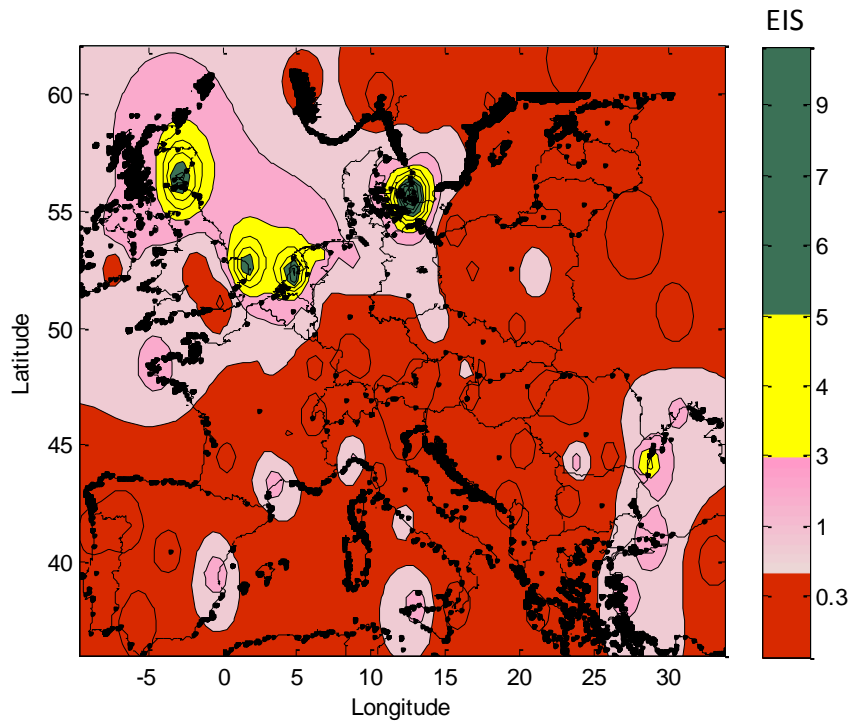

**Fig. 11.** Emergy Index of Sustainability, *EIS* (Installed power: 3000kW)

As it has been done for Copenhagen in Table 4, the *Emergy Index of Sustainability* (*EIS*) has been calculated for the 90 meteorological stations in Europe, for the three types of wind turbines. Thus, the maps proposed in Fig. 9 to Fig. 11 present the local sustainability of the system per unit of environmental loading, which can be a mean to identify the location offering the most sustainable potential for the implementation of wind turbines. Table 5 gives the details of the intermediary results: Wind kinetic energy, electricity produced, *EYR*, *ELR*, *EIS* and the transformity of the electricity produced. The different locations (cities) are ranked according to decreasing values of *EIS*. A first reading of this table then allows an identification of the most favorable places for wind turbines to be done (Copenhagen). It can also be noted that the decreasing values of *EIS* correspond to the decreasing values or *EYR* and increasing values of *ELR*; what could be expected because of the definitions of these indices. This order is not followed by the transformity, confirming that the transformity alone cannot be used to assess the sustainability of a system and/or availability of the resources and/or the environmental impact of a process. Brown et Ulgiati (2002) classified the energy sources according to the *EYR* value. For the authors, an *EYR*<2 means the energy source should be considered as a consumer of energy more than a source. This is the case for 70% of the locations for the 850kW wind turbines, 45% for the 1650kW wind turbines and 58% for the 3000kW wind turbines (Table 5, Fig. 9 to Fig.11). And for those having an *EYR*>2, they still must be considered has a secondary source more than a primary one, for which *EYR* should be highest than 5 (in Table 5, the highest value for *EYR* is 4.74, for the 1650kW wind turbine located at Copenhagen). For the same authors, an *EIS*<1 is indicative of products or processes that are not sustainable on the long run. They refine the analysis by classifying the products or processes having an *EIS* between 1 and 5 as medium run sustainable ones. In Table 5 it can be seen that only four locations offer a wind resource able to implemented 850kW wind turbines with an *EIS*>5. For the 3000kW wind turbines (which have the same wind speed characteristics as the 850kW ones and then are comparable each other) none of the locations offer a wind resource associated to an *EIS*>5.

The emergy indices chosen in this work are all dependant on the wind resources which are specific to each location. More precisely, as it is described in the appendix A, the produced electricity is function of the hourly difference between upstream and downstream wind speed. So, the electricity produced is dependent on the frequency distribution of wind speed for a given location. Fig. 12 shows this frequency distribution in a year for Hemsby, Copenhagen, Lyon, Bordeaux, Innsbruck and Venice. This helps to explain the ranking between these cities based on the sustainability potential of wind use for electricity production. In Table 5, Copenhagen is the most interesting place (with for example an *EIS*=8.57 and an *EYR*=3.47 for the 850kW wind turbine), before Hemsby (*EIS*=5.89 and *EYR*=2.98 for the same type of wind turbine), which is before Lyon (*EIS*=1.02 and *EYR*=1.62), etc. Indeed, even if Hemsby has a higher peak frequency of wind speed than Copenhagen, the latter has a wind resource which covers a larger spectrum of wind speeds. Lyon and Bordeaux also offer a large spectrum of useful wind speeds but with lower frequency. It is logical to have Innsbruck and Venice at the end of the ranking as the spectrum of useful wind speed is low and the high frequency showed in Fig. 12 corresponds to wind speeds lower than the cut-in wind speed.

Sensitivity analysis of design parameters of wind turbine (cut-in and nominal wind speeds) with *EIS* values are performed on six different areas for the 1650 kW wind turbine where *EIS* values are very large (Copenhagen and Hemsby), average (Lyon and Bordeaux) and small (Insbruck and Venice) as shown in Fig.13. In this figure, the cut-in wind speed variation ($\Delta V_{ci}$) is the difference between initial cut-in wind speed (3.5 m/s as given in technical specifications, Table 1) and increasing or decreasing value from it. Similarly, the nominal wind speed variation ($\Delta V_R$) is the difference between initial nominal wind speed (13 m/s as given in technical specifications, Table 1) and increasing or decreasing value from it. On the y-axis, *EIS* represents the *Emergy Index Sustainability* for initial cut-in and nominal wind speeds; and $\Delta EIS$ represents the difference between the initial *Emergy Index of Sustainability* values and the *Emergy Index of Sustainability* values obtained due to the increase or the decrease in cut-in and nominal wind speed. It is clear from Fig. 13 that *EIS* is more negatively sensitive than positively to the values of design parameters (cut-in and nominal wind speed). The figure further illustrates that the locations Copenhagen, Hemsby, Insbruck and Venice are less sensitive with the design parameter compare to Lyon and Bordeaux. This is due to the fact that wind speed distribution in Copenhagen and Hemsby are higher in contrast to Lyon and Bordeaux.

# 5 Conclusion

*Emergy Yield Ratio* (*EYR*) and *Emergy Index of Sustainability* (*EIS*) have been used to assess the potential of wind resources in 90 European regions. Due to their definitions these indices allowed a ranking to be done between the different locations, according to the sustainability of wind resources to produce electricity. The analysis of the results shows the combined effects of the wind turbine characteristics (cut-in and nominal wind speeds) and the local frequency distribution of wind speed on these indices. Coupled to the effect of wind turbine size (nominal power) it illustrates the importance of a local multicriterion analysis in the decision making of wind turbine implementation. The information given by the indicators of the emergy analysis helps to draw conclusions in this decision making process. Added to the information given by the transformity of the electricity produced (efficiency of the energy conversion), *EYR* and *EIS* indicate the potential of local resources contribution on economy and the sustainability of the system per unit of environmental loading, respectively. The results given by the use of these indices are in accordance with those provided by the literature (comparison of *EIS* and *EYR* for renewable resources use to produce electricity). They show that electricity from wind in most of the countries studied should be considered as a secondary resource. Moreover, most of the locations can be considered only as medium run sustainable ones. A non negligible number of the places are even considered as not sustainable on the long run. A sensitivity analysis of *EIS* to the wind turbine characteristics demonstrates the importance of the wind turbine type, in link with the local characteristics of the wind resource.

**Acknowledgement**
This research has been conducted in collaboration with Mines Nantes, Technische Universitet Eindhoven, funded through Erasmus Mundus Joint Doctoral Programme SELECT+, the support of which is gratefully acknowledged.

**Table 5 (Part a)** Emergy assessment indicators (90 cities, 3 types of wind turbines)

| Stations | 850 kW Wind Turbines | | | | | | | 1650 kW Wind Turbines | | | | | | | 3000 kW Wind Turbines | | | | | | |
|---|---|---|---|---|---|---|---|---|---|---|---|---|---|---|---|---|---|---|---|---|---|
| | Electrical output (J) | Kinetic energy (J) | R=(KE*τ) (seJ) | EYR | ELR | EIS | Transformity (seJ/J) | Electrical output (J) | Kinetic energy (J) | R=(KE*τ) (seJ) | EYR | ELR | EIS | Transformity (seJ/J) | Electrical output (J) | Kinetic energy (J) | R=(KE*τ) (seJ) | EYR | ELR | EIS | Transformity (seJ/J) |
| Copenhagen | 3,17E+14 | 2,62E+15 | 1,10E+19 | 3,47 | 0,40 | 8,57 | 4,88E+04 | 7,42E+14 | 7,44E+15 | 3,12E+19 | 4,74 | 0,27 | 17,76 | 4,06E+04 | 1,17E+15 | 1,00E+16 | 4,19E+19 | 4,02 | 0,33 | 12,15 | 4,78E+04 |
| Amsterdam | 2,52E+14 | 2,34E+15 | 9,83E+18 | 3,21 | 0,45 | 7,08 | 5,67E+04 | 6,03E+14 | 6,63E+15 | 2,78E+19 | 4,34 | 0,30 | 14,46 | 3,29E+04 | 9,30E+14 | 8,93E+15 | 3,75E+19 | 3,70 | 0,37 | 9,99 | 5,52E+04 |
| Leuchars | 2,47E+14 | 2,13E+15 | 8,92E+18 | 3,00 | 0,50 | 6,02 | 5,42E+04 | 5,96E+14 | 6,02E+15 | 2,52E+19 | 4,03 | 0,33 | 12,18 | 6,83E+04 | 9,18E+14 | 8,10E+15 | 3,40E+19 | 3,45 | 0,41 | 8,45 | 5,22E+04 |
| Hemsby | 2,75E+14 | 2,10E+15 | 8,80E+18 | 2,98 | 0,51 | 5,89 | 4,82E+04 | 6,72E+14 | 5,95E+15 | 2,49E+19 | 3,99 | 0,33 | 11,94 | 3,63E+04 | 1,03E+15 | 8,00E+15 | 3,36E+19 | 3,42 | 0,41 | 8,26 | 4,59E+04 |
| Constanta | 1,90E+14 | 1,70E+15 | 7,12E+18 | 2,60 | 0,63 | 4,16 | 6,10E+04 | 4,62E+14 | 4,80E+15 | 2,02E+19 | 3,42 | 0,41 | 8,26 | 4,07E+04 | 7,03E+14 | 6,47E+15 | 2,71E+19 | 2,96 | 0,51 | 5,78 | 5,83E+04 |
| Groningen | 2,24E+14 | 1,46E+15 | 6,14E+18 | 2,38 | 0,72 | 3,28 | 4,72E+04 | 5,67E+14 | 4,13E+15 | 1,73E+19 | 3,08 | 0,48 | 6,39 | 5,52E+04 | 8,58E+14 | 5,57E+15 | 2,34E+19 | 2,68 | 0,59 | 4,52 | 4,34E+04 |
| Istanbul | 2,48E+14 | 1,42E+15 | 5,95E+18 | 2,34 | 0,75 | 3,12 | 4,20E+04 | 6,20E+14 | 3,99E+15 | 1,67E+19 | 3,01 | 0,50 | 6,04 | 3,38E+04 | 9,56E+14 | 5,39E+15 | 2,26E+19 | 2,63 | 0,61 | 4,29 | 3,82E+04 |
| Brest | 2,31E+14 | 1,37E+15 | 5,76E+18 | 2,30 | 0,77 | 2,97 | 4,43E+04 | 5,87E+14 | 3,88E+15 | 1,63E+19 | 2,95 | 0,51 | 5,77 | 3,76E+04 | 8,92E+14 | 5,23E+15 | 2,19E+19 | 2,58 | 0,63 | 4,08 | 4,02E+04 |
| Palermo | 1,66E+14 | 1,36E+15 | 5,72E+18 | 2,29 | 0,78 | 2,94 | 6,12E+04 | 4,23E+14 | 3,84E+15 | 1,61E+19 | 2,93 | 0,52 | 5,67 | 5,70E+04 | 6,35E+14 | 5,19E+15 | 2,18E+19 | 2,57 | 0,64 | 4,03 | 5,61E+04 |
| Valencia | 1,37E+14 | 1,36E+15 | 5,69E+18 | 2,28 | 0,78 | 2,91 | 7,43E+04 | 3,61E+14 | 3,82E+15 | 1,60E+19 | 2,92 | 0,52 | 5,62 | 5,62E+04 | 5,30E+14 | 5,16E+15 | 2,17E+19 | 2,56 | 0,64 | 4,00 | 6,70E+04 |
| Montpellier | 2,04E+14 | 1,35E+15 | 5,65E+18 | 2,27 | 0,79 | 2,88 | 4,96E+04 | 5,27E+14 | 3,81E+15 | 1,60E+19 | 2,92 | 0,52 | 5,58 | 4,46E+04 | 7,86E+14 | 5,13E+15 | 2,15E+19 | 2,55 | 0,65 | 3,95 | 4,51E+04 |
| Brussel | 2,05E+14 | 1,33E+15 | 5,60E+18 | 2,26 | 0,79 | 2,84 | 4,90E+04 | 5,24E+14 | 3,76E+15 | 1,58E+19 | 2,89 | 0,53 | 5,48 | 4,61E+04 | 7,90E+14 | 5,08E+15 | 2,13E+19 | 2,54 | 0,65 | 3,89 | 4,46E+04 |
| Izmir | 2,17E+14 | 1,33E+15 | 5,59E+18 | 2,26 | 0,80 | 2,84 | 4,63E+04 | 5,41E+14 | 3,76E+15 | 1,58E+19 | 2,89 | 0,53 | 5,47 | 6,00E+04 | 8,33E+14 | 5,08E+15 | 2,13E+19 | 2,53 | 0,65 | 3,89 | 4,22E+04 |
| Craiova | 1,34E+14 | 1,32E+15 | 5,52E+18 | 2,24 | 0,81 | 2,78 | 7,42E+04 | 3,42E+14 | 3,72E+15 | 1,56E+19 | 2,87 | 0,53 | 5,38 | 1,05E+05 | 5,14E+14 | 5,01E+15 | 2,10E+19 | 2,51 | 0,66 | 3,81 | 6,79E+04 |
| Beek | 2,04E+14 | 1,32E+15 | 5,52E+18 | 2,24 | 0,81 | 2,78 | 4,88E+04 | 5,24E+14 | 3,70E+15 | 1,55E+19 | 2,86 | 0,54 | 5,33 | 8,31E+04 | 7,86E+14 | 5,00E+15 | 2,10E+19 | 2,51 | 0,66 | 3,80 | 4,44E+04 |
| Bremen | 2,07E+14 | 1,29E+15 | 5,40E+18 | 2,21 | 0,82 | 2,68 | 4,77E+04 | 5,37E+14 | 3,62E+15 | 1,52E+19 | 2,82 | 0,55 | 5,13 | 5,21E+04 | 8,02E+14 | 4,89E+15 | 2,05E+19 | 2,48 | 0,68 | 3,67 | 4,29E+04 |
| Clones | 1,95E+14 | 1,28E+15 | 5,36E+18 | 2,20 | 0,83 | 2,66 | 5,02E+04 | 5,00E+14 | 3,60E+15 | 1,51E+19 | 2,81 | 0,55 | 5,09 | 4,09E+04 | 7,50E+14 | 4,86E+15 | 2,04E+19 | 2,47 | 0,68 | 3,63 | 4,57E+04 |
| Dublin | 1,95E+14 | 1,28E+15 | 5,36E+18 | 2,20 | 0,83 | 2,66 | 5,02E+04 | 5,00E+14 | 3,60E+15 | 1,51E+19 | 2,81 | 0,55 | 5,09 | 6,18E+04 | 7,50E+14 | 4,86E+15 | 2,04E+19 | 2,47 | 0,68 | 3,63 | 4,57E+04 |
| Odessa | 2,19E+14 | 1,26E+15 | 5,28E+18 | 2,19 | 0,84 | 2,60 | 4,44E+04 | 5,68E+14 | 3,56E+15 | 1,49E+19 | 2,79 | 0,56 | 5,00 | 3,68E+04 | 8,57E+14 | 4,79E+15 | 2,01E+19 | 2,45 | 0,69 | 3,54 | 3,96E+04 |
| Genova | 1,88E+14 | 1,25E+15 | 5,23E+18 | 2,18 | 0,85 | 2,56 | 5,14E+04 | 4,63E+14 | 3,52E+15 | 1,48E+19 | 2,77 | 0,56 | 4,92 | 5,49E+04 | 7,18E+14 | 4,75E+15 | 1,99E+19 | 2,43 | 0,70 | 3,49 | 4,71E+04 |
| Vienna | 2,00E+14 | 1,23E+15 | 5,16E+18 | 2,16 | 0,86 | 2,50 | 4,81E+04 | 5,06E+14 | 3,46E+15 | 1,45E+19 | 2,74 | 0,57 | 4,78 | 3,70E+04 | 7,67E+14 | 4,68E+15 | 1,96E+19 | 2,41 | 0,71 | 3,41 | 4,37E+04 |
| Prague | 1,79E+14 | 1,21E+15 | 5,10E+18 | 2,15 | 0,87 | 2,46 | 5,33E+04 | 4,66E+14 | 3,43E+15 | 1,44E+19 | 2,73 | 0,58 | 4,71 | 5,66E+04 | 6,91E+14 | 4,62E+15 | 1,94E+19 | 2,40 | 0,72 | 3,35 | 4,81E+04 |
| Oban | 1,58E+14 | 1,16E+15 | 4,87E+18 | 2,09 | 0,91 | 2,29 | 5,88E+04 | 4,04E+14 | 3,26E+15 | 1,37E+19 | 2,64 | 0,61 | 4,34 | 5,58E+04 | 6,09E+14 | 4,41E+15 | 1,85E+19 | 2,33 | 0,75 | 3,11 | 5,32E+04 |
| Warsaw | 2,21E+14 | 1,13E+15 | 4,73E+18 | 2,06 | 0,94 | 2,19 | 4,15E+04 | 5,76E+14 | 3,19E+15 | 1,34E+19 | 2,61 | 0,62 | 4,19 | 3,97E+04 | 8,68E+14 | 4,28E+15 | 1,80E+19 | 2,29 | 0,77 | 2,97 | 3,67E+04 |
| Galati | 2,04E+14 | 1,10E+15 | 4,63E+18 | 2,04 | 0,96 | 2,12 | 4,44E+04 | 5,22E+14 | 3,12E+15 | 1,31E+19 | 2,57 | 0,64 | 4,04 | 6,01E+04 | 7,97E+14 | 4,19E+15 | 1,76E+19 | 2,27 | 0,79 | 2,87 | 3,95E+04 |
| Varna | 1,27E+14 | 1,07E+15 | 4,47E+18 | 2,01 | 0,99 | 2,02 | 7,02E+04 | 3,40E+14 | 3,01E+15 | 1,26E+19 | 2,51 | 0,66 | 3,81 | 5,70E+04 | 4,96E+14 | 4,06E+15 | 1,70E+19 | 2,23 | 0,82 | 2,73 | 6,23E+04 |
| Rome | 1,67E+14 | 1,06E+15 | 4,45E+18 | 2,00 | 1,00 | 2,00 | 5,33E+04 | 4,43E+14 | 2,98E+15 | 1,25E+19 | 2,50 | 0,67 | 3,75 | 5,00E+04 | 6,55E+14 | 4,03E+15 | 1,69E+19 | 2,22 | 0,82 | 2,71 | 4,70E+04 |
| Berlin | 1,99E+14 | 1,05E+15 | 4,39E+18 | 1,99 | 1,01 | 1,96 | 4,45E+04 | 5,28E+14 | 2,92E+15 | 1,23E+19 | 2,47 | 0,68 | 3,64 | 4,26E+04 | 7,81E+14 | 3,97E+15 | 1,67E+19 | 2,20 | 0,83 | 2,64 | 3,91E+04 |
| Dusseldorf | 1,86E+14 | 1,03E+15 | 4,33E+18 | 1,97 | 1,03 | 1,92 | 4,73E+04 | 4,86E+14 | 2,89E+15 | 1,21E+19 | 2,45 | 0,69 | 3,57 | 6,37E+04 | 7,24E+14 | 3,92E+15 | 1,64E+19 | 2,18 | 0,84 | 2,59 | 4,19E+04 |
| Goteborg | 1,92E+14 | 9,72E+14 | 4,08E+18 | 1,92 | 1,09 | 1,76 | 4,44E+04 | 5,16E+14 | 2,72E+15 | 1,14E+19 | 2,37 | 0,73 | 3,24 | 6,17E+04 | 7,55E+14 | 3,69E+15 | 1,55E+19 | 2,12 | 0,90 | 2,36 | 3,89E+04 |
| Thessaloniki | 1,31E+14 | 9,68E+14 | 4,06E+18 | 1,91 | 1,09 | 1,75 | 6,52E+04 | 3,44E+14 | 2,70E+15 | 1,13E+19 | 2,36 | 0,73 | 3,21 | 6,96E+04 | 5,09E+14 | 3,68E+15 | 1,54E+19 | 2,11 | 0,90 | 2,35 | 5,76E+04 |
| Hamburg | 1,93E+14 | 9,52E+14 | 4,00E+18 | 1,90 | 1,11 | 1,70 | 4,39E+04 | 5,08E+14 | 2,67E+15 | 1,12E+19 | 2,34 | 0,74 | 3,15 | 5,14E+04 | 7,56E+14 | 3,61E+15 | 1,52E+19 | 2,09 | 0,92 | 2,29 | 3,84E+04 |
| Paris | 1,83E+14 | 9,48E+14 | 3,98E+18 | 1,89 | 1,12 | 1,69 | 4,62E+04 | 4,97E+14 | 2,66E+15 | 1,12E+19 | 2,34 | 0,75 | 3,13 | 5,06E+04 | 7,22E+14 | 3,60E+15 | 1,51E+19 | 2,09 | 0,92 | 2,27 | 4,02E+04 |
| Saint-Hubert | 2,05E+14 | 9,17E+14 | 3,85E+18 | 1,86 | 1,16 | 1,61 | 4,05E+04 | 5,51E+14 | 2,57E+15 | 1,08E+19 | 2,29 | 0,77 | 2,96 | 5,73E+04 | 8,11E+14 | 3,48E+15 | 1,46E+19 | 2,05 | 0,95 | 2,16 | 3,51E+04 |
| Ostrava | 1,80E+14 | 8,97E+14 | 3,77E+18 | 1,85 | 1,18 | 1,56 | 4,56E+04 | 4,76E+14 | 2,53E+15 | 1,06E+19 | 2,27 | 0,79 | 2,90 | 5,96E+04 | 7,12E+14 | 3,41E+15 | 1,43E+19 | 2,03 | 0,97 | 2,09 | 3,96E+04 |
| Evora | 2,19E+14 | 8,95E+14 | 3,76E+18 | 1,84 | 1,18 | 1,56 | 3,76E+04 | 5,91E+14 | 2,51E+15 | 1,05E+19 | 2,26 | 0,79 | 2,86 | 3,57E+04 | 8,69E+14 | 3,39E+15 | 1,42E+19 | 2,03 | 0,97 | 2,08 | 3,24E+04 |
| Frankfurt | 1,69E+14 | 8,83E+14 | 3,71E+18 | 1,83 | 1,20 | 1,53 | 4,82E+04 | 4,48E+14 | 2,47E+15 | 1,04E+19 | 2,24 | 0,81 | 2,78 | 6,22E+04 | 6,62E+14 | 3,35E+15 | 1,41E+19 | 2,01 | 0,99 | 2,04 | 4,22E+04 |
| Kosice | 1,44E+14 | 8,72E+14 | 3,66E+18 | 1,82 | 1,22 | 1,50 | 5,65E+04 | 3,75E+14 | 2,46E+15 | 1,03E+19 | 2,24 | 0,81 | 2,77 | 5,60E+04 | 5,58E+14 | 3,32E+15 | 1,39E+19 | 2,00 | 1,00 | 2,01 | 4,98E+04 |
| Munich | 1,13E+14 | 8,42E+14 | 3,54E+18 | 1,79 | 1,26 | 1,43 | 7,07E+04 | 3,13E+14 | 2,34E+15 | 9,82E+18 | 2,18 | 0,85 | 2,57 | 6,61E+04 | 4,48E+14 | 3,20E+15 | 1,34E+19 | 1,97 | 1,03 | 1,90 | 6,10E+04 |
| Bergen | 1,46E+14 | 8,38E+14 | 3,52E+18 | 1,79 | 1,27 | 1,41 | 5,46E+04 | 3,81E+14 | 2,35E+15 | 9,85E+18 | 2,18 | 0,85 | 2,58 | 5,08E+04 | 5,69E+14 | 3,18E+15 | 1,34E+19 | 1,96 | 1,04 | 1,89 | 4,79E+04 |
| Brindisi | 1,58E+14 | 8,27E+14 | 3,47E+18 | 1,78 | 1,28 | 1,39 | 5,01E+04 | 4,19E+14 | 2,31E+15 | 9,69E+18 | 2,16 | 0,86 | 2,51 | 5,90E+04 | 6,22E+14 | 3,14E+15 | 1,32E+19 | 1,95 | 1,05 | 1,85 | 4,35E+04 |
| Nantes | 1,76E+14 | 7,97E+14 | 3,35E+18 | 1,75 | 1,33 | 1,32 | 4,43E+04 | 4,81E+14 | 2,24E+15 | 9,39E+18 | 2,13 | 0,89 | 2,39 | 4,44E+04 | 7,04E+14 | 3,02E+15 | 1,27E+19 | 1,91 | 1,09 | 1,75 | 3,77E+04 |
| Kilkenny | 1,46E+14 | 7,66E+14 | 3,22E+18 | 1,72 | 1,38 | 1,24 | 5,26E+04 | 3,88E+14 | 2,14E+15 | 8,98E+18 | 2,08 | 0,93 | 2,24 | 7,95E+04 | 5,72E+14 | 2,91E+15 | 1,22E+19 | 1,88 | 1,14 | 1,65 | 4,56E+04 |
| Palma | 1,47E+14 | 7,63E+14 | 3,21E+18 | 1,72 | 1,39 | 1,24 | 5,21E+04 | 3,82E+14 | 2,14E+15 | 8,99E+18 | 2,08 | 0,93 | 2,24 | 5,82E+04 | 5,76E+14 | 2,90E+15 | 1,22E+19 | 1,88 | 1,14 | 1,65 | 4,52E+04 |
| Plovdiv | 1,06E+14 | 7,62E+14 | 3,20E+18 | 1,72 | 1,39 | 1,24 | 7,20E+04 | 2,73E+14 | 2,15E+15 | 9,02E+18 | 2,08 | 0,92 | 2,25 | 6,85E+04 | 4,06E+14 | 2,90E+15 | 1,22E+19 | 1,88 | 1,14 | 1,65 | 6,41E+04 |
| Helsinki | 1,78E+14 | 7,58E+14 | 3,18E+18 | 1,72 | 1,40 | 1,23 | 4,29E+04 | 4,86E+14 | 2,11E+15 | 8,87E+18 | 2,06 | 0,94 | 2,20 | 9,67E+04 | 7,09E+14 | 2,87E+15 | 1,21E+19 | 1,87 | 1,15 | 1,62 | 3,66E+04 |
| Faro | 1,60E+14 | 7,53E+14 | 3,16E+18 | 1,71 | 1,41 | 1,22 | 4,76E+04 | 4,35E+14 | 2,09E+15 | 8,76E+18 | 2,05 | 0,95 | 2,16 | 6,04E+04 | 6,38E+14 | 2,85E+15 | 1,20E+19 | 1,86 | 1,16 | 1,61 | 4,06E+04 |
| Birmingham | 1,78E+14 | 7,42E+14 | 3,12E+18 | 1,70 | 1,43 | 1,19 | 4,25E+04 | 4,89E+14 | 2,05E+15 | 8,62E+18 | 2,03 | 0,97 | 2,10 | 6,45E+04 | 7,11E+14 | 2,81E+15 | 1,18E+19 | 1,85 | 1,18 | 1,57 | 3,61E+04 |

**Table 5 (Part b)** Emergy assessment indicators (90 cities, 3 types of wind turbines)

| | 850 kW Wind Turbines | | | | | | | 1650 kW Wind Turbines | | | | | | | 3000 kW Wind Turbines | | | | | | |
|---|---|---|---|---|---|---|---|---|---|---|---|---|---|---|---|---|---|---|---|---|---|
| Stations | Electrical output (J) | Kinetic energy (J) | R=(KE*τ) (seJ) | EYR | ELR | EIS | Transformity (seJ/J) | Electrical output (J) | Kinetic energy (J) | R=(KE*τ) (seJ) | EYR | ELR | EIS | Transformity (seJ/J) | Electrical output (J) | Kinetic energy (J) | R=(KE*τ) (seJ) | EYR | ELR | EIS | Transformity (seJ/J) |
| Naples | 9,72E+13 | 7,18E+14 | 3,02E+18 | 1,68 | 1,48 | 1,14 | 7,68E+04 | 2,73E+14 | 2,00E+15 | 8,39E+18 | 2,01 | 0,99 | 2,02 | 7,43E+04 | 3,90E+14 | 2,73E+15 | 1,14E+19 | 1,82 | 1,21 | 1,50 | 6,49E+04 |
| Athens | 1,53E+14 | 6,97E+14 | 2,93E+18 | 1,66 | 1,52 | 1,09 | 4,83E+04 | 4,11E+14 | 1,95E+15 | 8,20E+18 | 1,98 | 1,02 | 1,95 | 5,54E+04 | 6,06E+14 | 2,64E+15 | 1,11E+19 | 1,80 | 1,25 | 1,44 | 4,12E+04 |
| Karlstad | 1,72E+14 | 6,74E+14 | 2,83E+18 | 1,64 | 1,57 | 1,04 | 4,24E+04 | 4,73E+14 | 1,87E+15 | 7,87E+18 | 1,94 | 1,06 | 1,84 | 5,74E+04 | 6,88E+14 | 2,55E+15 | 1,07E+19 | 1,77 | 1,30 | 1,37 | 3,57E+04 |
| Santander | 1,16E+14 | 6,74E+14 | 2,83E+18 | 1,64 | 1,57 | 1,04 | 6,29E+04 | 3,17E+14 | 1,87E+15 | 7,87E+18 | 1,94 | 1,06 | 1,84 | 7,80E+04 | 4,59E+14 | 2,56E+15 | 1,07E+19 | 1,77 | 1,29 | 1,37 | 5,36E+04 |
| Lyon | 1,27E+14 | 6,62E+14 | 2,78E+18 | 1,62 | 1,60 | 1,02 | 5,70E+04 | 3,53E+14 | 1,85E+15 | 7,75E+18 | 1,93 | 1,08 | 1,79 | 9,08E+04 | 5,09E+14 | 2,51E+15 | 1,05E+19 | 1,76 | 1,32 | 1,34 | 4,80E+04 |
| Barcelona | 1,50E+14 | 6,56E+14 | 2,76E+18 | 1,62 | 1,62 | 1,00 | 4,80E+04 | 4,24E+14 | 1,84E+15 | 7,71E+18 | 1,92 | 1,08 | 1,78 | 5,25E+04 | 6,06E+14 | 2,48E+15 | 1,04E+19 | 1,75 | 1,33 | 1,31 | 4,01E+04 |
| Stockholm | 1,37E+14 | 6,36E+14 | 2,67E+18 | 1,60 | 1,67 | 0,96 | 5,19E+04 | 3,95E+14 | 1,75E+15 | 7,36E+18 | 1,88 | 1,13 | 1,66 | 5,99E+04 | 5,57E+14 | 2,40E+15 | 1,01E+19 | 1,73 | 1,37 | 1,26 | 4,30E+04 |
| Poznan | 1,38E+14 | 6,04E+14 | 2,54E+18 | 1,57 | 1,75 | 0,89 | 5,06E+04 | 3,93E+14 | 1,69E+15 | 7,10E+18 | 1,85 | 1,17 | 1,58 | 5,67E+04 | 5,59E+14 | 2,28E+15 | 9,59E+18 | 1,69 | 1,45 | 1,17 | 4,20E+04 |
| Andravida | 1,20E+14 | 5,95E+14 | 2,50E+18 | 1,56 | 1,78 | 0,88 | 5,77E+04 | 3,29E+14 | 1,67E+15 | 7,00E+18 | 1,84 | 1,19 | 1,54 | 6,57E+04 | 4,80E+14 | 2,26E+15 | 9,48E+18 | 1,68 | 1,46 | 1,15 | 4,87E+04 |
| Belgrade | 1,21E+14 | 5,95E+14 | 2,50E+18 | 1,56 | 1,78 | 0,88 | 5,76E+04 | 3,40E+14 | 1,65E+15 | 6,95E+18 | 1,83 | 1,20 | 1,53 | 8,47E+04 | 4,86E+14 | 2,25E+15 | 9,46E+18 | 1,68 | 1,47 | 1,15 | 4,80E+04 |
| Kohn | 1,38E+14 | 5,89E+14 | 2,47E+18 | 1,56 | 1,80 | 0,86 | 5,00E+04 | 3,90E+14 | 1,62E+15 | 6,80E+18 | 1,82 | 1,23 | 1,48 | 6,92E+04 | 5,58E+14 | 2,23E+15 | 9,35E+18 | 1,67 | 1,48 | 1,13 | 4,16E+04 |
| London | 1,38E+14 | 5,86E+14 | 2,46E+18 | 1,55 | 1,81 | 0,86 | 5,01E+04 | 3,86E+14 | 1,61E+15 | 6,78E+18 | 1,81 | 1,23 | 1,47 | 1,06E+05 | 5,56E+14 | 2,22E+15 | 9,31E+18 | 1,67 | 1,49 | 1,12 | 4,17E+04 |
| Linz | 1,15E+14 | 5,83E+14 | 2,45E+18 | 1,55 | 1,82 | 0,85 | 6,01E+04 | 3,13E+14 | 1,62E+15 | 6,80E+18 | 1,81 | 1,23 | 1,48 | 5,89E+04 | 4,56E+14 | 2,21E+15 | 9,29E+18 | 1,67 | 1,49 | 1,12 | 5,08E+04 |
| Strasbourg | 1,25E+14 | 5,73E+14 | 2,41E+18 | 1,54 | 1,85 | 0,83 | 5,51E+04 | 3,51E+14 | 1,60E+15 | 6,74E+18 | 1,81 | 1,24 | 1,46 | 7,68E+04 | 5,02E+14 | 2,17E+15 | 9,11E+18 | 1,66 | 1,52 | 1,09 | 4,58E+04 |
| Porto | 1,20E+14 | 5,65E+14 | 2,37E+18 | 1,53 | 1,88 | 0,82 | 5,68E+04 | 3,32E+14 | 1,55E+15 | 6,51E+18 | 1,78 | 1,28 | 1,39 | 7,39E+04 | 4,80E+14 | 2,14E+15 | 8,98E+18 | 1,65 | 1,55 | 1,07 | 4,76E+04 |
| Krakow | 1,22E+14 | 5,55E+14 | 2,33E+18 | 1,52 | 1,91 | 0,80 | 5,57E+04 | 3,30E+14 | 1,56E+15 | 6,54E+18 | 1,78 | 1,28 | 1,40 | 6,15E+04 | 4,84E+14 | 2,10E+15 | 8,83E+18 | 1,64 | 1,57 | 1,04 | 4,69E+04 |
| Madrid | 1,01E+14 | 5,52E+14 | 2,32E+18 | 1,52 | 1,92 | 0,79 | 6,70E+04 | 2,78E+14 | 1,54E+15 | 6,48E+18 | 1,78 | 1,29 | 1,38 | 8,44E+04 | 4,03E+14 | 2,09E+15 | 8,80E+18 | 1,63 | 1,58 | 1,04 | 5,62E+04 |
| Pisa | 1,12E+14 | 5,49E+14 | 2,31E+18 | 1,52 | 1,93 | 0,79 | 6,02E+04 | 3,10E+14 | 1,52E+15 | 6,39E+18 | 1,77 | 1,30 | 1,35 | 1,03E+05 | 4,49E+14 | 2,08E+15 | 8,75E+18 | 1,63 | 1,59 | 1,03 | 5,04E+04 |
| Bordeaux | 1,37E+14 | 5,42E+14 | 2,28E+18 | 1,51 | 1,95 | 0,77 | 4,90E+04 | 3,95E+14 | 1,51E+15 | 6,32E+18 | 1,76 | 1,32 | 1,33 | 6,71E+04 | 5,59E+14 | 2,05E+15 | 8,60E+18 | 1,62 | 1,61 | 1,00 | 4,02E+04 |
| Stuttgart | 1,11E+14 | 5,09E+14 | 2,14E+18 | 1,48 | 2,08 | 0,71 | 5,94E+04 | 3,11E+14 | 1,40E+15 | 5,88E+18 | 1,71 | 1,42 | 1,20 | 8,24E+04 | 4,45E+14 | 1,93E+15 | 8,10E+18 | 1,58 | 1,71 | 0,92 | 4,94E+04 |
| Podgorica | 9,09E+13 | 5,01E+14 | 2,11E+18 | 1,47 | 2,11 | 0,70 | 7,21E+04 | 2,58E+14 | 1,40E+15 | 5,88E+18 | 1,70 | 1,42 | 1,20 | 8,81E+04 | 3,66E+14 | 1,90E+15 | 7,97E+18 | 1,57 | 1,74 | 0,90 | 5,97E+04 |
| Nancy | 1,22E+14 | 4,83E+14 | 2,03E+18 | 1,46 | 2,19 | 0,66 | 5,32E+04 | 3,51E+14 | 1,34E+15 | 5,63E+18 | 1,67 | 1,48 | 1,13 | 9,03E+04 | 4,96E+14 | 1,82E+15 | 7,67E+18 | 1,55 | 1,81 | 0,86 | 4,34E+04 |
| Braganca | 9,28E+13 | 4,64E+14 | 1,95E+18 | 1,44 | 2,28 | 0,63 | 6,89E+04 | 2,65E+14 | 1,28E+15 | 5,37E+18 | 1,64 | 1,55 | 1,06 | 1,10E+05 | 3,75E+14 | 1,75E+15 | 7,37E+18 | 1,53 | 1,88 | 0,81 | 5,66E+04 |
| Bratislava | 1,09E+14 | 4,48E+14 | 1,89E+18 | 1,42 | 2,36 | 0,60 | 5,80E+04 | 3,18E+14 | 1,25E+15 | 5,25E+18 | 1,63 | 1,59 | 1,03 | 6,81E+04 | 4,46E+14 | 1,69E+15 | 7,12E+18 | 1,51 | 1,95 | 0,78 | 4,71E+04 |
| Ankara | 1,04E+14 | 4,45E+14 | 1,87E+18 | 1,42 | 2,38 | 0,60 | 6,11E+04 | 2,88E+14 | 1,23E+15 | 5,19E+18 | 1,62 | 1,61 | 1,01 | 6,86E+04 | 4,14E+14 | 1,68E+15 | 7,08E+18 | 1,51 | 1,96 | 0,77 | 5,06E+04 |
| Kiev | 1,12E+14 | 4,40E+14 | 1,85E+18 | 1,42 | 2,40 | 0,59 | 5,65E+04 | 3,37E+14 | 1,21E+15 | 5,11E+18 | 1,61 | 1,63 | 0,99 | 7,74E+04 | 4,62E+14 | 1,66E+15 | 6,97E+18 | 1,50 | 1,99 | 0,75 | 4,51E+04 |
| Mannheim | 1,22E+14 | 4,33E+14 | 1,82E+18 | 1,41 | 2,45 | 0,58 | 5,14E+04 | 3,58E+14 | 1,17E+15 | 4,94E+18 | 1,59 | 1,69 | 0,94 | 6,30E+04 | 5,01E+14 | 1,63E+15 | 6,85E+18 | 1,49 | 2,03 | 0,74 | 4,14E+04 |
| Debrecen | 9,10E+13 | 4,03E+14 | 1,70E+18 | 1,38 | 2,62 | 0,53 | 6,76E+04 | 2,66E+14 | 1,12E+15 | 4,70E+18 | 1,56 | 1,77 | 0,88 | 1,02E+05 | 3,73E+14 | 1,52E+15 | 6,40E+18 | 1,46 | 2,17 | 0,67 | 5,44E+04 |
| Szombathely | 9,10E+13 | 4,03E+14 | 1,70E+18 | 1,38 | 2,62 | 0,53 | 6,76E+04 | 2,66E+14 | 1,12E+15 | 4,70E+18 | 1,56 | 1,77 | 0,88 | 7,70E+04 | 3,73E+14 | 1,52E+15 | 6,40E+18 | 1,46 | 2,17 | 0,67 | 5,44E+04 |
| Messina | 9,75E+13 | 4,02E+14 | 1,69E+18 | 1,38 | 2,63 | 0,52 | 6,30E+04 | 2,91E+14 | 1,08E+15 | 4,56E+18 | 1,55 | 1,83 | 0,85 | 1,05E+05 | 4,02E+14 | 1,52E+15 | 6,37E+18 | 1,46 | 2,18 | 0,67 | 5,04E+04 |
| Sevilla | 1,06E+14 | 3,95E+14 | 1,66E+18 | 1,37 | 2,68 | 0,51 | 5,77E+04 | 3,11E+14 | 1,08E+15 | 4,53E+18 | 1,54 | 1,84 | 0,84 | 7,60E+04 | 4,36E+14 | 1,49E+15 | 6,25E+18 | 1,45 | 2,22 | 0,65 | 4,62E+04 |
| Kolobrzeg | 1,13E+14 | 3,93E+14 | 1,65E+18 | 1,37 | 2,69 | 0,51 | 5,38E+04 | 3,37E+14 | 1,08E+15 | 4,56E+18 | 1,55 | 1,83 | 0,84 | 5,78E+04 | 4,68E+14 | 1,48E+15 | 6,22E+18 | 1,45 | 2,23 | 0,65 | 4,29E+04 |
| Bucharest | 7,59E+13 | 3,82E+14 | 1,61E+18 | 1,36 | 2,77 | 0,49 | 7,98E+04 | 2,20E+14 | 1,07E+15 | 4,49E+18 | 1,54 | 1,86 | 0,83 | 1,50E+05 | 3,11E+14 | 1,45E+15 | 6,08E+18 | 1,44 | 2,28 | 0,63 | 6,42E+04 |
| Tampere | 1,05E+14 | 3,60E+14 | 1,52E+18 | 1,34 | 2,93 | 0,46 | 5,66E+04 | 3,18E+14 | 9,70E+14 | 4,08E+18 | 1,49 | 2,04 | 0,73 | 7,74E+04 | 4,37E+14 | 1,35E+15 | 5,69E+18 | 1,41 | 2,44 | 0,58 | 4,48E+04 |
| Geneva | 7,75E+13 | 3,43E+14 | 1,44E+18 | 1,32 | 3,08 | 0,43 | 7,61E+04 | 2,17E+14 | 9,46E+14 | 3,98E+18 | 1,48 | 2,09 | 0,71 | 1,07E+05 | 3,12E+14 | 1,30E+15 | 5,46E+18 | 1,39 | 2,54 | 0,55 | 6,20E+04 |
| Oslo | 8,28E+13 | 3,25E+14 | 1,37E+18 | 1,31 | 3,25 | 0,40 | 7,03E+04 | 2,39E+14 | 8,87E+14 | 3,73E+18 | 1,45 | 2,23 | 0,65 | 1,07E+05 | 3,36E+14 | 1,23E+15 | 5,16E+18 | 1,37 | 2,69 | 0,51 | 5,67E+04 |
| Venice | 1,06E+14 | 3,17E+14 | 1,33E+18 | 1,30 | 3,34 | 0,39 | 9,57E+04 | 1,79E+14 | 8,61E+14 | 3,62E+18 | 1,43 | 2,30 | 0,62 | 1,58E+05 | 2,46E+14 | 1,20E+15 | 5,03E+18 | 1,36 | 2,76 | 0,49 | 7,69E+04 |
| Sofia | 7,47E+13 | 3,09E+14 | 1,30E+18 | 1,29 | 3,42 | 0,38 | 7,70E+04 | 2,16E+14 | 8,60E+14 | 3,62E+18 | 1,43 | 2,30 | 0,62 | 1,15E+05 | 3,05E+14 | 1,17E+15 | 4,91E+18 | 1,35 | 2,82 | 0,48 | 6,17E+04 |
| Salzburg | 8,49E+13 | 2,67E+14 | 1,13E+18 | 1,25 | 3,95 | 0,32 | 6,57E+04 | 2,63E+14 | 7,00E+14 | 2,95E+18 | 1,35 | 2,83 | 0,48 | 8,58E+04 | 3,56E+14 | 1,00E+15 | 4,21E+18 | 1,30 | 3,29 | 0,40 | 5,08E+04 |
| Cluj | 5,55E+13 | 2,62E+14 | 1,10E+18 | 1,25 | 4,03 | 0,31 | 1,00E+05 | 1,67E+14 | 7,23E+14 | 3,04E+18 | 1,37 | 2,74 | 0,50 | 1,21E+05 | 2,30E+14 | 9,90E+14 | 4,17E+18 | 1,30 | 3,33 | 0,39 | 7,84E+04 |
| Minsk | 8,87E+13 | 2,57E+14 | 1,08E+18 | 1,24 | 4,10 | 0,30 | 6,24E+04 | 2,83E+14 | 6,95E+14 | 2,93E+18 | 1,35 | 2,85 | 0,47 | 1,07E+05 | 3,77E+14 | 9,60E+14 | 4,04E+18 | 1,29 | 3,43 | 0,38 | 4,76E+04 |
| Insbruck | 5,54E+13 | 2,33E+14 | 9,83E+17 | 1,22 | 4,53 | 0,27 | 9,81E+04 | 1,59E+14 | 6,37E+14 | 2,69E+18 | 1,32 | 3,11 | 0,43 | 6,43E+04 | 2,25E+14 | 8,81E+14 | 3,71E+18 | 1,27 | 3,74 | 0,34 | 7,83E+04 |

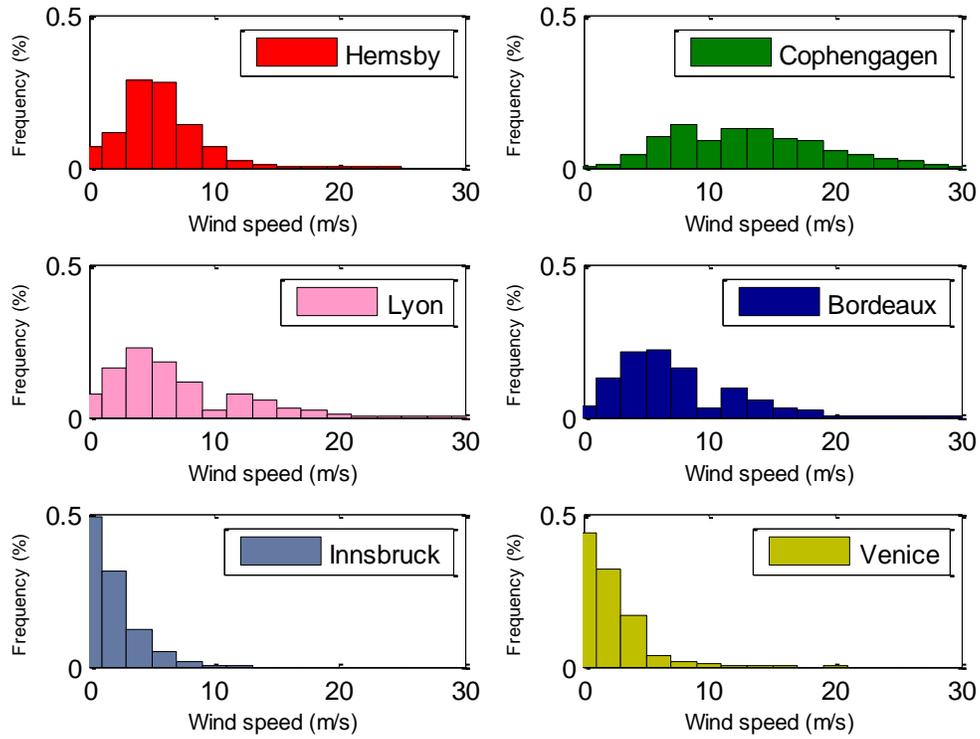

**Fig. 12** Frequency distribution of wind speed (Hemsby, Cophenhagen, Lyon, Bordeaux, Insbruck and Venice)

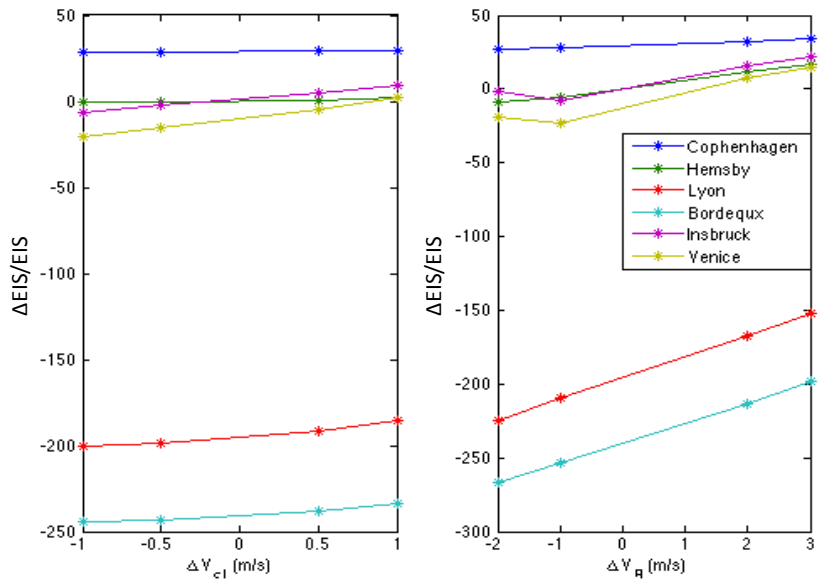

**Fig. 13.** Sensitivity analysis for change in start-up wind speed ($\Delta V_{ci}$) and change in nominal wind speed ($\Delta V_R$) with respect to ratio of deviation of change in emergy sustainability index to *EIS* values ($\frac{\Delta EIS}{EIS}$)

**Appendix A**

The power generated from wind turbine depends on hub height, available wind velocity and performance characteristics of wind generator. The wind speed at specific hub height of wind turbine can be estimated from wind speed measured at reference height (measurement height of wind speed) and is given in Hellman equation, see Eq.1, where $V$ is wind speed at hub height $z$ and $V_0$ is wind speed at reference height $z_0$.

$$V = V_0 \left(\frac{z}{z_0}\right)^\alpha \quad (1)$$

In Eq. 1, $\alpha$ represents the wind speed adjustment factor (Hellman coefficient of 0.28 see Ozgener et Ozgener, 2007).

Kinetic energy available to wind turbine ($KE_{wind}$) can be expressed in terms of air mass ($m$) and average wind speed ($V$).

$$KE_{wind} = \frac{1}{2} mV^2 \quad (2)$$

Kinetic energy can also be expressed in terms of density i.e. mass per unit volume (Eq. 3). Air volume crossing the wind turbine can be calculated as the product of cross-sectional area ($A$) perpendicular to direction of wind by horizontal length ($L$) of incoming wind, where, $= V.t$ , $t$ is the time and $\rho_a$ is density of air.

$$KE_{wind} = \frac{1}{2} \rho_a A t V^3 \quad (3)$$

Thus, power available to wind turbine is the time rate of kinetic energy, $\dot{KE}_{wind}$, and can be expressed in Eq. 4 (Dincer et al., 2007).

$$\dot{KE}_{wind} = \frac{1}{2} \rho_a A V^3 \quad (4)$$

Given $\dot{m}$ as the mass flow rate, rate of momentum change equals to the power available to wind turbine. Power absorbed from wind turbine (P) can be expressed in Eq. 5 as:

$$P = \dot{m}(V_1 - V_2) \bar{V} \quad (5)$$

$V_1$ and $V_2$ in Eq. 5 represents upstream and downstream wind speed and $\bar{V}$ is the wind speed available in the middle part of wind turbine system. Similarly, the rate of change of kinetic energy, $\dot{K}_W$ can be expressed as:

$$\dot{K}_W = \frac{1}{2} \dot{m}(V_1^2 - V_2^2) \quad (6)$$

Eq. 5 and Eq. 6 are equal so that retardation of the wind $V_1 - \bar{V}$, before the rotor of wind turbine is equal to retardation $\bar{V} - V_2$, with the assumption that direction of wind velocity through rotor of wind turbine is axial and velocity is uniform over the area (Dincer et al., 2007). Rate of kinetic energy from the wind turbine is given in Eq. 7, where, $\bar{V} = \frac{V_1+V_2}{2}$ .

$$\dot{KE}_W = \rho A \bar{V}(V_1 - V_2)\bar{V} = \rho A \left(\frac{V_1 + V_2}{2}\right)^2 (V_1 - V_2) \quad (7)$$

For the calculation of the rate of kinetic energy from incoming wind, downstream wind speed $V_2$ is neglected from the difference of upstream and downstream wind speed in Eq. 7 as $V_2$ is available at the output of wind turbine. However, average wind speed is available in wind turbine system. So, Eq. 7 can be further deduced from Eq. 8 for calculation of the rate of kinetic energy available from incoming wind ($\dot{\psi}_{wind}$).

$$\dot\psi_{wind} = \frac{1}{2}\rho_a A V_1^2 \left(\frac{V_1 + V_2}{2}\right) \qquad (8)$$

Electrical output of wind turbines depends on the performance characteristics of the wind generator. When wind speed is below start-up wind speed, the wind turbine will not be able to generate power. When wind speed is above start-up wind speed, the wind turbine produces power nonlinearly and increases till it reaches to rated nominal power for increasing wind speed to the maximum value. Power output from the wind turbine is zero after cut-out wind speed. Based on these assumptions, linear (Bueno et al., 2005), weibull parameter (Borowy et al., 1996) and quadratic (Lu et al., 2002; Chou et al., 1981) model can be found in the literature. For this study, simplified Chou wind turbine model is considered since this model is generalized for any wind turbine system, which considers start-up and rated wind speed and is given in Eq. 9, where, $P_{WT}$ represents power available from wind turbine.

$$P_{WT} = \begin{cases} 0 & V \le V_{ci}; V > V_{co} \\ A + BV + CV^2 & V_{ci} < V \le V_R \\ P_R & V_R < V \le V_{co} \end{cases} \qquad (9)$$

The variables $A$, $B$, and $C$ can be chosen based on requirement of linear or quadratic model, and for this study, variables A, B and C are chosen for the linear range operation. In Eq. 9, $V_{ci}$ is start-up (cut-in) wind speed, $V_R$ is rated (nominal) wind speed, $V_{co}$ is cut-out wind speed and $P_R$ is rated power of the wind turbine.

Wind height adjustment, $V$, is calculated based on Eq. 1 for incoming wind speed, $V_0$, where, speed adjustment factor $\alpha$ is 0.28. Downstream wind speed $V_2$ is calculated from iterative procedure based on power generated from wind turbine and initial wind speed distribution. Kinetic energy available at each station is calculated by summing rate of change of kinetic energy of each hourly data of a year (8760 data/year). Electrical power is then calculated for one year. Fig. A1 shows electrical power for each station for 850 kW, 1650 kW and 3000 kW wind turbines at different wind speeds. It is clear that power generation from wind turbine is zero until it reaches 3.5 m/s for 1650 kW wind turbines and 4 m/s for 850 kW and 3000 kW wind turbines. Hence the power goes linearly till 13.5 m/s for 1650 kW wind turbines and 16 m/s for 850 kW and 3000 kW wind turbines. It can be noted that Chou's model is similar to Pedersen model (Pedersen et al., 1992), which is linear by parts. After 13.5 m/s (1650 kW wind turbines) and 16 m/s (850 kW and 3000 kW wind turbines) till 25 m/s of wind speed, the power equal to the full rated power of wind turbines and becomes zero.

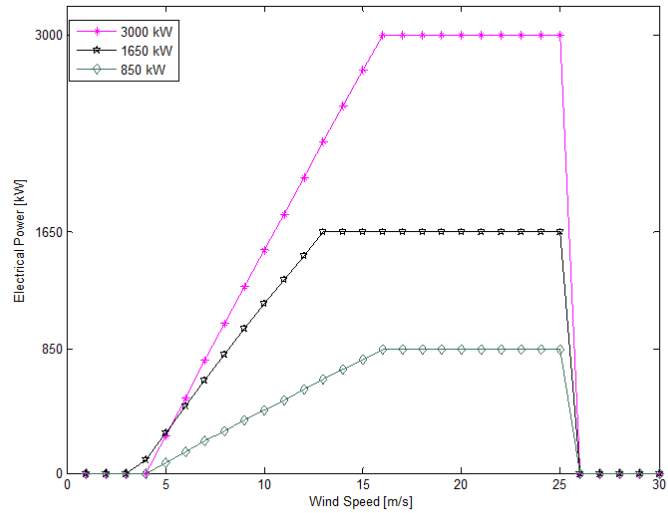

**Fig. A1.** Electrical power generation from 850 kW, 1650 kW and 3000 kW wind turbines

Variation of $V_1$ with change in wind speed ($\Delta V = V_1 - V_2$) is shown in Fig. A2. It is clear that there is no change in wind speed when $V_1$ is below 4 m/s and above 25 m/s, but its variation is high when $V_1$ is between 4 to 6 m/s. This variation is high since less wind resource is available which results to minimum power production from wind turbine. Also, it is clear that variation of wind speed decreases linearly with increasing of $V_1$ beyond 6 m/s. This is because the wind turbine is able to produce electrical power linearly after this velocity. Fig. A3 shows the electrical power generation from 850 kW, 1650 kW and 3000 kW wind turbines for change in upstream and downstream wind speed $\Delta V$. It delineates that when $\Delta V$ is 0.5 – 1.6 m/s in 850 kW and 3000 kW wind turbines and 0.5 – 2 m/s in 1650 kW wind turbines, power generated from the wind turbine is linearly increased and also remains constant to the wind turbine nominal power.

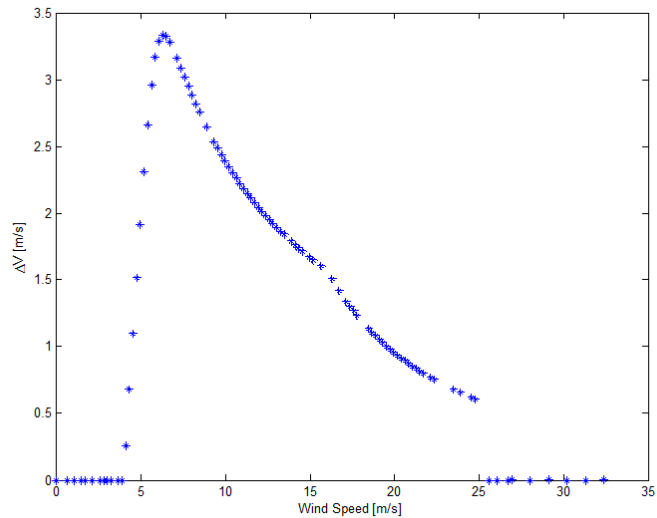

**Fig. A2.** Change in wind speed with respect to upstream wind

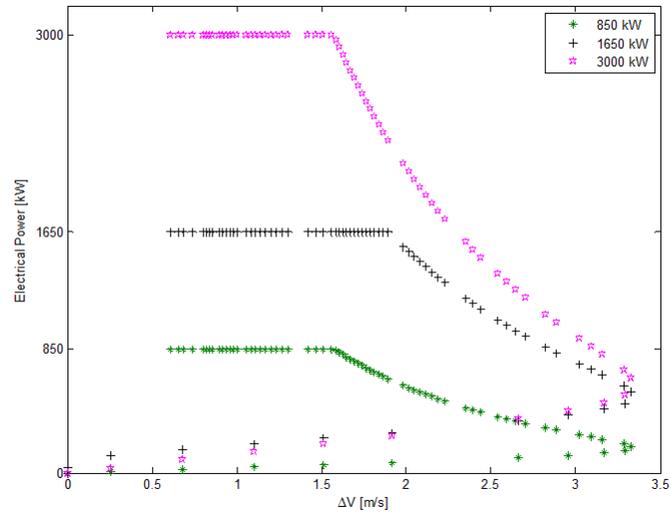

**Fig. A3.** Electrical power generation with respect to change in upstream and downstream wind speed from 850 kW, 1650 kW and 3000 kW wind turbines